\documentclass[12pt,a4paper]{scrartcl}
\usepackage[utf8]{inputenc}
\usepackage{amsmath}
\usepackage{graphicx}
\usepackage{amssymb}
\usepackage{authblk}
\usepackage{color}
\definecolor{sienna}{rgb}{0.53, 0.18, 0.09}
\definecolor{tealgreen}{rgb}{0.0, 0.51, 0.5}
\definecolor{rossocorsa}{rgb}{0.83, 0.0, 0.0}
\usepackage[sort]{cite}
\usepackage{hyperref}
\addtokomafont{caption}{\small}
\addtokomafont{captionlabel}{\bfseries}
\setcapindent{0pt}
\usepackage{multirow}
\usepackage{makecell}

\newcommand{\hl}{}
\DeclareMathOperator{\re}{Re}
\title{Thermal instabilities, frequency comb formation, and temporal oscillations in Kerr microresonators}
\author[1]{Amir Leshem}
\affil[1]{Racah Institute of Physics, The Hebrew University, Jerusalem, Israel 9190401}
\author[2]{Zhen Qi}
\author[2]{Thomas F. Carruthers}
\author[2]{Curtis R.\ Menyuk}
\affil[2]{University of Maryland at Baltimore County, 1000 Hilltop Circle, Baltimore, MD 21250, USA}
\author[1]{Omri Gat}
\begin{document}

\maketitle
\begin{abstract}
We analyze the consequences of dissipative heating in driven Kerr microresonators theoretically and numerically, using a thermal Lugiato-Lefever model. We show that thermal sensitivity modifies the stability range of continuous wave in a way that blocks direct access to broadband frequency-comb forming waveforms, and we propose a deterministic access path that bypasses the thermal instability barrier. We describe a novel thermal instability that leads to thermooptical oscillations via a Hopf bifurcation.
\end{abstract}
\section{Introduction}
The great progress in the generation of frequency combs in driven Kerr microresonators \cite{Pfeiffer:2017ia,Spencer:2018eb,Kippenberg:2018hi,Chembo:2015hz,Pasquazi:2018de,Karpov:2019cs} has motivated a flurry of theoretical effort to identify and calculate waveforms in microresonator models \cite{Chembo:2010cb,MatskoAB:2011ws,Chembo:2013ew,ParraRivas:2014kk,Godey:2014cj,Bao:2017jz,Godey:2017kt,Perinet:2017ct,ParraRivas:2018ie,Hansson:2018ie,Qi:2017da,Qi:2019fla,Kholmyansky:2019cy,Milian:2015db,Kartashov:2017cj}. A thorough mapping of the steady states of microresonator models and their stability plays an important role in the experimental search for deterministically accessible waveforms with good comb properties.

The steady state and stability properties of continuous waves, cnoidal waves (also known as perfect soliton crystals, Turing rolls and periodic patterns \cite{Qi:2019fla}), and solitons in the basic Lugiato-Lefever model of microresonators are quite well-understood \cite{Barashenkov:1996te,MatskoAB:2011ws,Qi:2017da,Qi:2019fla,Kholmyansky:2019cy}, and significant progress has been made in the study of multisoliton waveforms \cite{Barashenkov:1998wo,ParraRivas:2018ie}, and of solitons with high-order dispersion and Raman scattering \cite{Milian:2015db,Bao:2017jz}. 

However, this theoretical progress has for the most part neglected the role of dissipative heating on the resonator waveforms and their stability, although the thermal response plays a crucial part in the experimental generation of cavity solitons.  The cooling of the resonator when the power circulating in the cavity drops abruptly after a chaotic waveform evolves into a bunch of solitons creates a pump-cavity detuning drift, potentially destabilizing the solitons. \hl{The cooling instability is an obstacle to soliton formation that can be mitigated with thermal control \cite{joshiThermallyControlledComb2016}, or by operating at cryogenic temperatures \cite{moilleKerrMicroresonatorSolitonFrequency2019}}. On the other hand, the temperature dependence of the cavity resonance separates the overlapping stability regions of soliton and multisoliton waveforms \cite{ADeterministicMeth:2020tk}, facilitating the synthesis of a single-soliton waveforms by backward detuning \cite{Guo:2016es}, and can serve as a means to control the dynamics of cavity solitons \cite{Stone:2018jr}.

Dissipative heating plays an important role in applications of Kerr microresonators other than  frequency comb formation, and this effect was studied in numerous experiments, where it often leads to temporal oscillations of a spatially uniform field; see \cite{Jiang:2020ia} for a review of experiments and potential applications. While this phenomenon has been studied extensively in experiments and simulations, a basic theoretical understanding of the onset of oscillations and its dependence on parameters is presently lacking.

In this paper we apply the dynamical computation approach \cite{Wang:2014ki} to the problem of thermooptical dynamics of Kerr resonators. This method, which combines efficient computational tools with ideas from dynamical systems theory, allows for a comprehensive mapping of stable solutions of nonlinear evolution equations for large sets of parameter values. 
\hl{In previous works we applied this method to models of mode-locked lasers \cite{Wang:2014ki,Wang:2016jz,Wang:2016un,Qi:2017da,Wang:2018gp} and pumped Kerr resonators \cite{Kholmyansky:2019cy,Qi:2019fla}, where we studied the existence and stability of dissipative cnoidal waves. Cnoidal waves are spatially periodic waveforms that can arise as Turing patterns by modulational instability of continuous-wave light for blue- and moderately red-detuned pumping, but become a periodic train of well-separated pulses in the highly red-detuned pump regime, where stable cnoidal waves coexist with stable single cavity solitons. This regime also supports stationary nonperiodic arrays of solitons, such as soliton crystals with defects \cite{coleSolitonCrystalsKerr2017}, and accordingly highly red-detuned cnoidal waves are also known as perfect soliton crystals \cite{Karpov:2019cs,hePerfectSolitonCrystals2020}. Here the term cnoidal waves will refer to any periodic waveform, of which perfect soliton crystals is the highly red-detuned subset.}

\hl{In applications of Kerr microresonators to frequency comb formation the goal is to generate strong coherent wide band combs \cite{Pfeiffer:2017ia,Spencer:2018eb,Kippenberg:2018hi,Chembo:2015hz,Pasquazi:2018de,Karpov:2019cs}. The current standard practice for this purpose is to use single soliton waveforms as a source. However, single solitons are not directly accessible from continuous-wave light, and are instead generated by sweeping the pump frequency through chaotic regions, a process which yields cavity solitons only with some probability. Another limitation of single solitons as comb sources is that they use the pump inefficiently and yield weak combs, especially for relatively large resonators that have radio-frequency soliton repetition rates \cite{yiSolitonFrequencyComb2015}.}

\hl{Cnoidal waves on the other hand are deterministically accessible and can have bandwidths comparable with solitons. Of these, the red-detuned perfect soliton crystals exhibit the best comb shape and utilize the pump power most efficiently, and these waveforms can be accessed by paths in parameter space that were identified using dynamical computational methods in \cite{Kholmyansky:2019cy,Qi:2019fla}.} It should be emphasized that the waveform evolution along paths of this type, which moves through stable steady states, is completely deterministic in nature, depending neither on the rate of path traversal, as long as it slow enough that the evolution is adiabatic, nor on initial conditions, in contrast with stochastic control paths that reach soliton crystals with some probability from chaotic waveforms \cite{Karpov:2019cs}.

Here we compute the existence and stability properties of continuous waves including thermal effects, and we explain the similarities and differences from the well-studied case in which thermal effects are ignored \cite{ParraRivas:2014kk,Godey:2014cj}. We identify three basic consequences of dissipative heating: First, as the thermal sensitivity increases, the coexistence wedge in parameter space, where there are three continuous wave solutions for each set of pump parameters, moves down in pump power; as a result, thermal instabilities block access paths which start with a moderately or highly red-detuned pump frequency. We show in Fig.\ \ref{fig:alphaf} examples of the coexistence wedge and an access path; the significance of the parameters is explained below. \hl{The thermal shift of the nonlinear fold of the resonance curve and of the bistability region was identified and studied in detail in \cite{carmonDynamicalThermalBehavior2004}}. 

Second, the threshold curve of modulational instability, where cnoidal waves bifurcate from continuous waves, moves to larger detuning. Therefore, a deterministic path to wide-band perfect soliton crystals  must \emph{bypass} the tip of the coexistence wedge, before crossing the lower end of the modulational instability curve. 

\hl{The two thermal effects described so far depend only on the strength of the thermal response and are insensitive to the thermal time scale. For this reason they have close analogues in Fabry-Perot Kerr resonators, where interactions between counter-propagating pulses gives rise to a detuning proportional to the total power like the one caused by changing temperature, albeit on a completely different time scale \cite{obrzudTemporalSolitonsMicroresonators2017,coleTheoryKerrFrequency2018}.}

The third consequence of dissipative heating is that continuous waves become susceptible to instabilities that lead to the aforementioned oscillations of spatially uniform waveforms. We show that oscillatory instabilities arise only when the heating rate is larger than a minimal threshold value. Then continuous waves become unstable either for sufficiently large pump powers for any detuning, or only for a range of pump powers above a minimal detuning, depending on the thermal response parameters in a way which is explicitly derived.

It has been argued that the competition between processes, such as temperature dependence of the index of refraction versus thermal expansion, is the common mechanism in the several physical underpinnings of thermal oscillations \cite{Jiang:2020ia}. An interesting example is the thermal relaxation oscillations that occur because of competition of thermal detuning and thermal expansion  \cite{He:2009fv,Diallo:2015ck}, where theoretical modeling has to track the temperatures of the mode volume and the bulk of the microresonator separately. In other cases \cite{Fomin:2005vv,Park:2007wf} the thermal oscillations arise through thermal detuning alone, which depends only on the mode volume temperature, which is included in standard models of thermal response of comb-forming microresonators (e.g.\ \cite{Guo:2016es}), and this approach will be adopted here.

After presenting the equations of motion governing the cavity waveform and thermal detuning in section \ref{sec:model}, we derive the basic properties and stability equations of stationary solutions in section \ref{sec:sta}. In section \ref{sec:cw} we calculate the continuous wave solutions and present their three modes of instability. Modulational instabilities can lead to the formation of steady cnoidal waves that are studied in section \ref{sec:cnoidal}, and the thermal oscillations that can arise from uniform instabilities are investigated in section \ref{sec:osc}. Section \ref{sec:conclusions} presents our conclusions.

\section{Thermooptical equations of motion}\label{sec:model}
We model the evolution of the slowly varying envelope $\psi(x,t)$ of the optical waveform in a pumped resonator with second order dispersion coefficient $\beta$ and Kerr coefficient $\gamma$ by the Lugiato-Lefever equation \hl{with periodic boundary conditions}
\begin{equation}
\frac{\partial\psi}{\partial t}=\bigl\{-l+i[\omega-\omega_c(T)]\bigr\}\psi-\frac{i\beta}{2}\frac{\partial^2\psi}{\partial x^2}+i\gamma|\psi|^2\psi+\tilde F\ ,\quad \hl{\psi(x=0,t)=\psi(x=\tilde L,t)\ ,}
\end{equation}
\hl{where $0\le x\le \tilde L$ is the position along the propagation direction in the resonator, $\tilde L$ is the resonant mode diameter,} $l$ is the loss coefficient, $\omega$ and $\tilde F$ are the pump frequency and amplitude (respectively), and $\omega_c$ is the temperature-dependent cavity resonance frequency \cite{baoDirectSolitonGeneration2017}.

The dynamics of the temperature $T$ of the resonator is governed by absorptive heating proportional to the light intensity, and diffusive cooling. The cooling rate is such that the typical time scale of temperature dynamics is always much longer than the cavity roundtrip time; therefore $T$ is independent of $x$, and dissipative heating is proportional to the mean power
\begin{equation}
\hl{P=\int_0^{\tilde L}|\psi|^2\frac{dx}{\tilde L}\ .}
\end{equation}
The heat balance equation therefore becomes
\begin{equation}
\frac{dT}{dt}=\lambda P-\kappa(T-T_0)\ ,
\end{equation}
where the energy conversion factor $\lambda$ and the cooling rate $\kappa$ are constant parameters that depend on material properties and the geometry of the resonator, and $T_0$ is the temperature of the environment. 
The dependence of the refractive index on temperature is weak so that we can express the thermal detuning as 
\begin{equation}
\Theta\equiv\omega_c(T)-\omega_c(T_0)=\omega_c'(T_0)(T-T_0)\ .
\end{equation}

Assuming anomalous dispersion $\beta<0$, and choosing units of time, space, and power such that $l=|\beta|=\gamma=1$ brings the equations of motion to dimensionless form
\begin{align}
\frac{\partial\psi}{\partial t}&=-\bigl(1+i(\alpha+\Theta)\bigr)\psi+ \frac{i}{2}\frac{\partial^2\psi}{\partial x^2}+i|\psi|^2\psi+F\ ,\quad \hl{\psi(x=0,t)=\psi(x= L,t)\ ,}\label{eq:ll}\\
\frac{d\Theta}{dt}&=-AP-B\Theta\ ,\label{eq:th}
\end{align}
where 
\begin{equation}
\hl{P=\int_0^{L}|\psi|^2\frac{dx}{L}\ ,}
\end{equation}
and the dimensionless parameters are
\begin{align}
\alpha&=\frac{\omega-\omega_c(T_0)}{l}\ ,\qquad F=\frac{\sqrt\gamma\tilde F}{l^{3/2}}\ ,\qquad L=\sqrt{\frac{l}{\beta}}\tilde L\ ,\\A&=-\frac{\lambda l^3}{\gamma\omega_c'}\ , \qquad \qquad\;B=\kappa l\ . \label{eq:thpar}
\end{align}
\hl{$\alpha$ and $F$ have the usual significance of pump detuning and amplitude, $L$ is the resonant mode length measured in units determined by the dispersion and loss, while $A$ is the heat-detuning conversion coefficient and $B$ is the thermal relaxation coefficient of the resonator.} In microcomb experiments $A$ is positive because $\omega_c$ is a decreasing function of temperature, and $B$ is positive by fundamental principles. \hl{As a consequence the thermal detuning $\Theta$ is always negative in the steady state and for resonators initially at ambient temperature.} The parameters $A$ and $B$ are typically small, of $O(10^{-1})$ or less \cite{ADeterministicMeth:2020tk}, but the thermal sensitivity ratio $C=A/B$ is of order one or larger, which means that thermal effects are moderate or strong, as we show below.

\section{Stationary solutions and their stability}\label{sec:sta}
If the cavity waveform $\psi_s(x)$ with mean power $P_s$ is independent of time, then the thermal detuning is
\begin{equation}
\Theta_s=-CP_s\ ,
\end{equation}
where the thermal sensitivity parameter $C=A/B$ as defined above. It follows that $\psi_s(x)$ obeys the integrodifferential equation
\begin{equation}\label{eq:stationary}
-\bigl[1+i(\alpha-CP_s)\bigr]\psi_s+ \frac{i}{2}\frac{\partial^2\psi_s}{\partial x^2}+i|\psi_s|^2\psi_s+F=0\ .
\end{equation}

A simple but important conclusion is that the entire set of stationary waveforms for \emph{all} detunings is independent of $C$. That is, if $\psi_s(\alpha,F,C;x)$ is a solution of Eq.\ \eqref{eq:stationary} with pump parameters $\alpha$, $F$, and thermal sensitivity ratio $C$, then the same waveform is a solution of this equation \emph{without} thermal response, lower detuning $\alpha_0=\alpha-CP_s$, and the same $F$:
\begin{equation}\label{eq:smap}
\psi_s(\alpha,F,C;x)= \psi_s(\alpha_0,F,0;x)\ ,\ \ \alpha=\alpha_0+CP_s\ .
\end{equation}
Regions of existence of stationary waves are accordingly shifted to larger detunings in the $\alpha$-$F$ plane as $C$ increases. Note however that although the correspondence \eqref{eq:smap} between stationary waveforms with different thermal coefficients is one-to-one, there are usually several solutions of Eq.\ \eqref{eq:stationary} for each choice of the parameters $\alpha$, $F$, $C$, with different mean powers, and these correspond via Eq.\ \eqref{eq:smap} to $C=0$ solutions with different values of $\alpha_0$, and vice versa. For example, the mean power of red-detuned cnoidal waves is much larger than that of single solitons with the same pump parameters, so that their thermal detuning is correspondingly larger.
\hl{As shown in \cite{obrzudTemporalSolitonsMicroresonators2017,coleTheoryKerrFrequency2018}, Eq.\ \eqref{eq:stationary}  with an appropriate choice of $C$ serves as an effective model for stationary waveforms in a Fabry-Perot Kerr resonator and therefore the mapping \eqref{eq:smap} also relates stationary waveform in ring and Fabry-Perot resonators.}

Note also that Eq.\ \eqref{eq:smap} does not imply that the waveform has the same stability properties for different thermal parameters. Stability of $\psi_s$ is determined in the standard manner by setting $\psi=\psi_s+\psi_1$ and $\Theta=-CP_s+\Theta_1$, with $\psi_1,\,\Theta_1$ small, in  \eqref{eq:ll}--\eqref{eq:th}, giving
\begin{align}
\frac{\partial\psi_1}{\partial t}&=-\bigl[1+i(\alpha-CP_s)\bigr]\psi_1-i\Theta_1\psi_s+ \frac{i}{2}\frac{\partial^2\psi_1}{\partial x^2}+2i|\psi_s|^2\psi_1+i\psi_s^2\psi_1^*\ ,\label{eq:ll1}\\
\frac{d\Theta_1}{dt}&=-AP_1-B\Theta_1\ ,\label{eq:th1}
\end{align}
where
\begin{equation}
P_1=2\re\int \frac{dx}{L}\psi_s(x)^*\psi_1(x)\ .
\end{equation}
In particular, the stability properties depend on the individual values of $A$ and $B$, and not only on their ratio  $C$.

\section{Instabilities of continuous waves}\label{sec:cw}
Stationary continuous waves are uniform solutions $\psi_s(x)\equiv\psi_c$ of Eq.\ \eqref{eq:stationary}; the mean power in this case is simply $|\psi_c|^2$, so that the thermal nonlinearity can be combined with the Kerr nonlinearity, giving
\begin{equation}\label{eq:cw}
-\bigl(1+i\alpha\bigr)\psi_c+i(1+C)|\psi_c|^2\psi_c+F=0\ .
\end{equation}
The thermal nonlinearity combines with the Kerr nonlinearity in Eq.\ \eqref{eq:cw} because for continuous waves, in contrast with other waveforms, the disparity of the time scales of the two nonlinearities does not matter.
It follows that for continuous waves there is a second mapping between thermal and nonthermal waveforms: If $\psi_c(\alpha,F,C)$ is a solution of Eq.\ \eqref{eq:cw} with pump parameters $\alpha,F$ and thermal sensitivity $C$, then $\sqrt{1+C}\psi_c(\alpha,F,C)$ is a solution for the same $\alpha$, $F_0=\sqrt{1+C}F$, and $C=0$,
\begin{equation}\label{eq:cmap}
\psi_c(\alpha,F,C)=\sqrt{1+C}\psi_c(\alpha,F_0,0)\ ,\qquad F_0=\sqrt{1+C}F\ .
\end{equation}
 The mapping \eqref{eq:cmap} relates different waveforms at the same pump frequency and different pump powers, whereas the mapping \eqref{eq:smap} relates the same waveform at the same pump power but at a different frequency.

It can be shown (see e.g.\ Ref.\ \cite{Godey:2014cj}) that Eq.\ \eqref{eq:cw} implies that the continuous wave power $|\psi_c|^2$ obeys a cubic equation. The real roots of the equation are always positive, and each such root corresponds to a unique complex solution of Eq.\ \eqref{eq:cw}. It follows that for a fixed choice of parameters, there are either one or three continuous wave solutions. When $C=0$ the region of coexistence of three continuous wave solutions is a wedge in parameter space, whose small-detuning tip is at $\alpha=\sqrt3$ (see Fig. \ref{fig:alphaf}).
The mapping \eqref{eq:cmap} implies that as $C$ is increased, the coexistence region in the $\alpha$-$F$ plane, 
\begin{equation}
F_{\text{ce}-}(\alpha)< F<{F_{\text{ce}+}(\alpha)}\ ,\qquad \alpha>\sqrt3\ ,
\end{equation}
 scales {down} in $F$ as $1/\sqrt{1+C}$ . Thus, 
the coexistence regions are wedge-shaped domains for any $C$, that are shown  in Fig.\ \ref{fig:alphaf} with blue boundaries. The expressions for $F_{\text{ce}\pm}$ are collected, along with definitions for additional quantities defined below, in Table \ref{tab:defs}. 

\begin{table}[t]
\changefontsizes{11pt}
\begin{tabular}{c|c|c|c}
Value & Expression & {Significance} & Graphical presentation\\[1mm]\hline\\[-2mm]
$F_{\text{ce}\pm}$ & $\sqrt{\displaystyle\frac{2}{27}\frac{\alpha(\alpha^2+9)\pm(\alpha^2-3)^{3/2}}{1+C}}$ & \multirow{2}{*}{\makecell{Boundaries of the\\ coexistence region}} & \multirow{2}{*}{Continuous blue line}\\[5mm]
$P_{\text{ce}\pm}$ & $\displaystyle\frac{2\alpha\pm2\sqrt{\alpha^2-3}}{3(1+C)}$ &  & \\[5mm]
$P_{\text{br}\pm}$ & $\displaystyle\frac{2\alpha\pm\sqrt{\alpha^2-3}}{3(1+C)}$ & \makecell{Boundaries between\\ cont.\ wave branches} & Dashed blue line\\[5mm]
$F_\text{mi}$&$\sqrt{1+(\alpha-1-C)^2}$&\makecell{Modulational insta-\\bility threshold}&Yellow line\\[5mm]
$\alpha_t$ & $2(1+C)-\sqrt{C(2+C)}$&\makecell{End of modulational \\instability curve}& Small black circle\\[5mm]
$P_{\text{lw}\pm}$&$\displaystyle\frac{(2+C)\alpha\pm\sqrt{\alpha^2-(C+1)(C+3)}}{(C+1)(C+3)}$&\makecell{Boundaries of long-\\wave instability region}& Red line\\[5mm]
$\alpha_h$&$\displaystyle\frac{2 (B+1) \sqrt{(C+1) (3+C-A)}}{A-2}$&\makecell{Minimal $\alpha$ for finite-\\frequency instability}& Tip of green line\\[5mm]
$P_{h\pm}$& $\displaystyle\frac{\alpha\bigl(2(2+C)-A\bigr)\pm\sqrt{\Delta}}{2(1+C)(3+C-A)}$&\multirow{2}{*}{\makecell{Boundaries of finite-\\frequency instability\\region}}& \multirow{2}{*}{Green line}\\[5mm]
$\Delta$&$\begin{array}{c}\alpha^2(A-2)^2-4(B+1)^2\\\qquad\times(3+C-A)(1+C)\end{array}$
\end{tabular}
\caption{\label{tab:defs} Analytical expressions and physical significance of the boundaries of the regions of existence and stability for continuous waves in Kerr resonators taking into account thermal effects (see main text for detailed description.) The boundaries and thresholds are shown in Figs.\ \ref{fig:alphaf} and \ref{fig:alphap}.}
\end{table}

It is useful to study the behavior of continuous waves as a function of $\alpha$ and $P$, because unlike the $\alpha$-$F$ parametrization, each choice of $\alpha$ and $P$ corresponds to a unique solution. The coexistence wedge in the $\alpha$-$F$ plane unfolds onto a smooth lobe $P_{\text{ce}-}<P<P_{\text{ce}+}$ in the $\alpha$-$P$ plane (see Table \ref{tab:defs}) bounded by the full blue curve in Fig.\ \ref{fig:alphap}. Since each point in the $\alpha$-$P$ plane corresponds to a single continuous wave, the three continuous wave branches appear as separate regions in the lobe bounded by $P_{\text{ce}\pm}$; the boundaries between these regions are the curves $P_{\text{br}\pm}(\alpha)$ (see Table \ref{tab:defs}) that are marked by the blue dashed curves in Fig.\ \ref{fig:alphap}, so that the continuous wave solutions in the regions below, between, and above the blue dashed lines belong to the lower, middle, and upper branches (respectively). The mapping \eqref{eq:cmap} implies that the lobe of coexistence scales down in $P$ as $1/(1+C)$ when $C$ is increased.

\subsection{Modulational instability}\label{sec:mod}
We next study the instability modes of \hl{continuous waves in} microresonators, taking into account thermal effects. Due to translational invariance, Eq.\ \eqref{eq:ll1} has solutions of the form $\psi_1(x,t)=\psi_{k+}(t)e^{ikx}+\psi_{k-}(t)e^{-ikx}$, $k$ real, for uniform $\psi_s\equiv\psi_c$. The thermal effects are nonlocal, so that it is necessary to consider the cases of zero and nonzero $k$ separately. In this subsection we focus on the $k\ne0$ case.

In the case $k\ne0$ we find that $P_1=0$, and Eq.\ \eqref{eq:th1} then implies that $\Theta_1=0$. Eq.\ \eqref{eq:ll1} becomes a \hl{two-variable} linear system in the variables $\psi_{k+}$, $\psi_{k-}$,
\begin{align}\label{eq:mod}
&\frac{d}{dt}\begin{pmatrix}\psi_{k+}\\\psi_{k-}\end{pmatrix}=(-1+iM_2)\begin{pmatrix}\psi_{k+}\\\psi_{k-}\end{pmatrix}\\
&M_2=
\begin{bmatrix}-\bigl(\alpha+\frac{1}{2}k^2-({2+C})|\psi_c|^2\bigr)&\psi_c^2\\-(\psi_c^*)^2&\alpha+\frac{1}{2}k^2-({2+C})|\psi_c|^2\end{bmatrix}
\end{align}
The eigenvalues of $iM_2$ are
\begin{equation}\label{eq:detm2}
\pm \sqrt{\det M_2}\ ,\qquad \det M_2=P^2-\bigl[k^2+\alpha-({2+C})P\bigr]^2\ ,
\end{equation}
so that the solution is modulationally unstable if $\max_{k} \det M_2>1$. 

\begin{figure}[p]
\centering
\centering\large\qquad$C=0$\qquad\qquad\qquad\qquad\qquad\qquad$C=5,\,B=0.05$\\
\includegraphics[width=7cm]{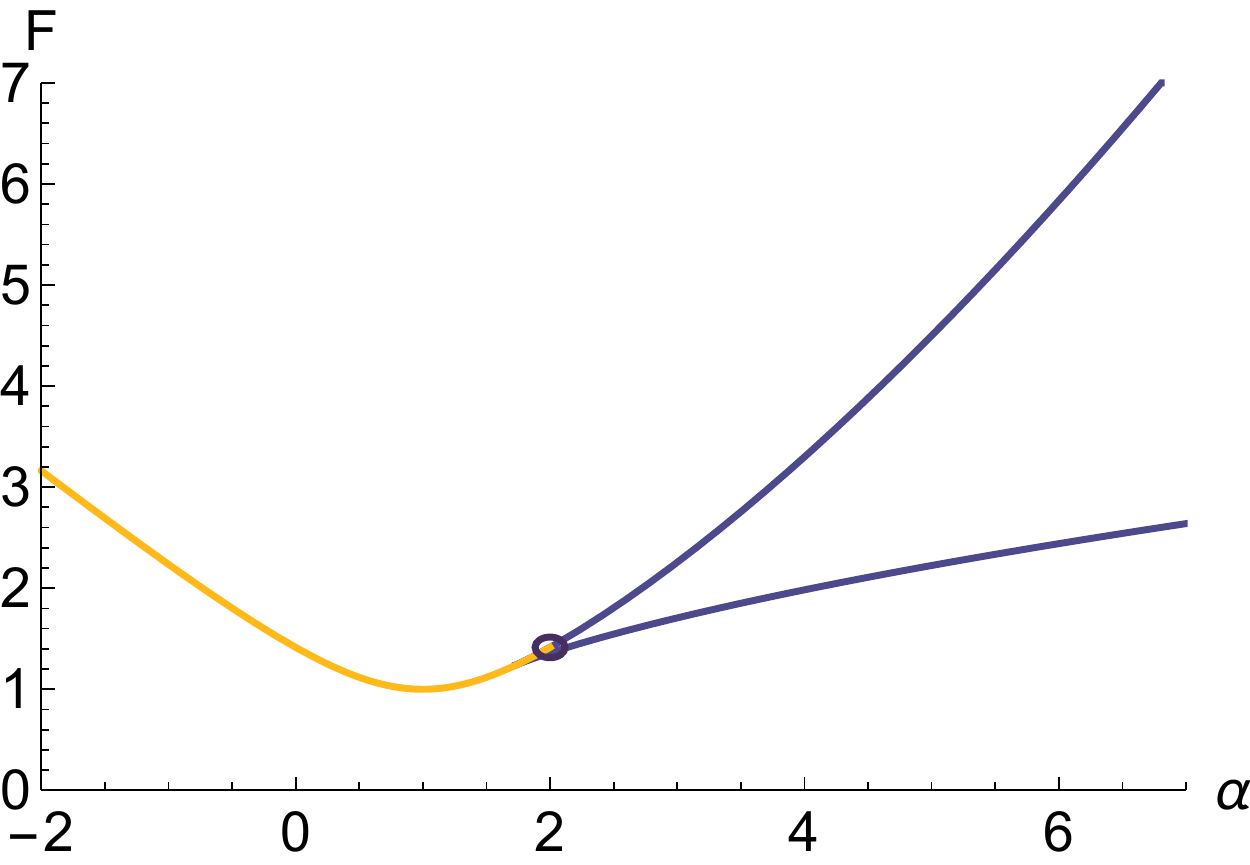}\quad\includegraphics[width=7cm]{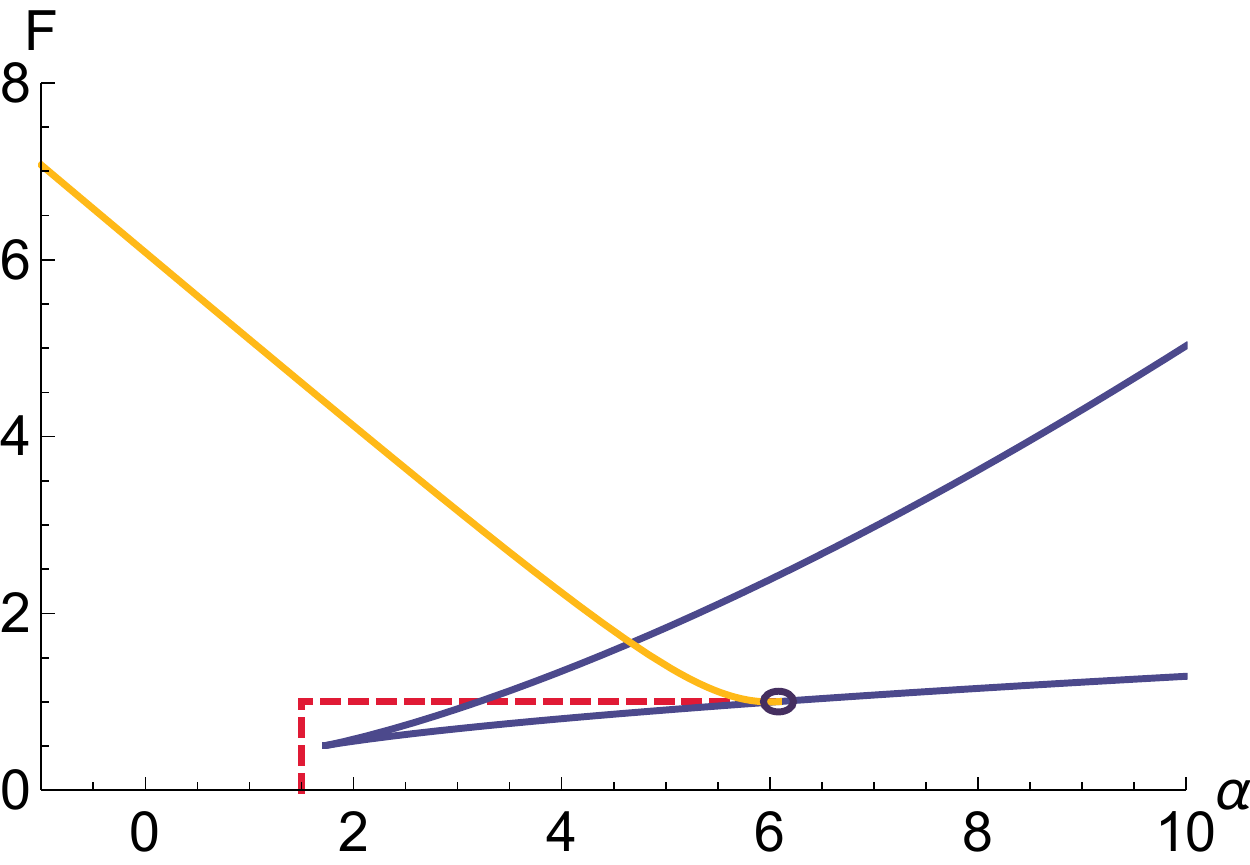}\\[5mm]\qquad$C=5,\,B=2$\qquad\qquad\qquad\qquad\quad$C=10,\,B=0.6$\\
\includegraphics[width=7cm]{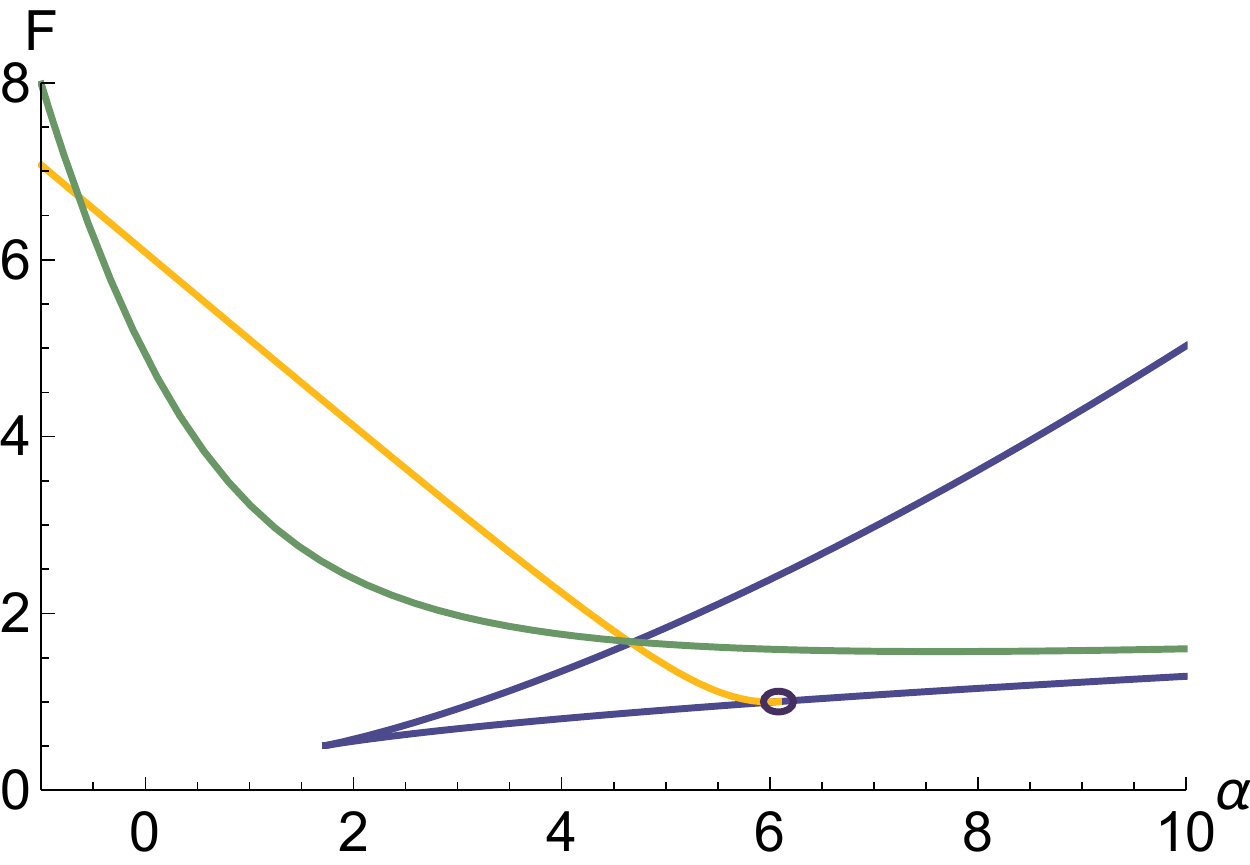}\quad\includegraphics[width=7cm]{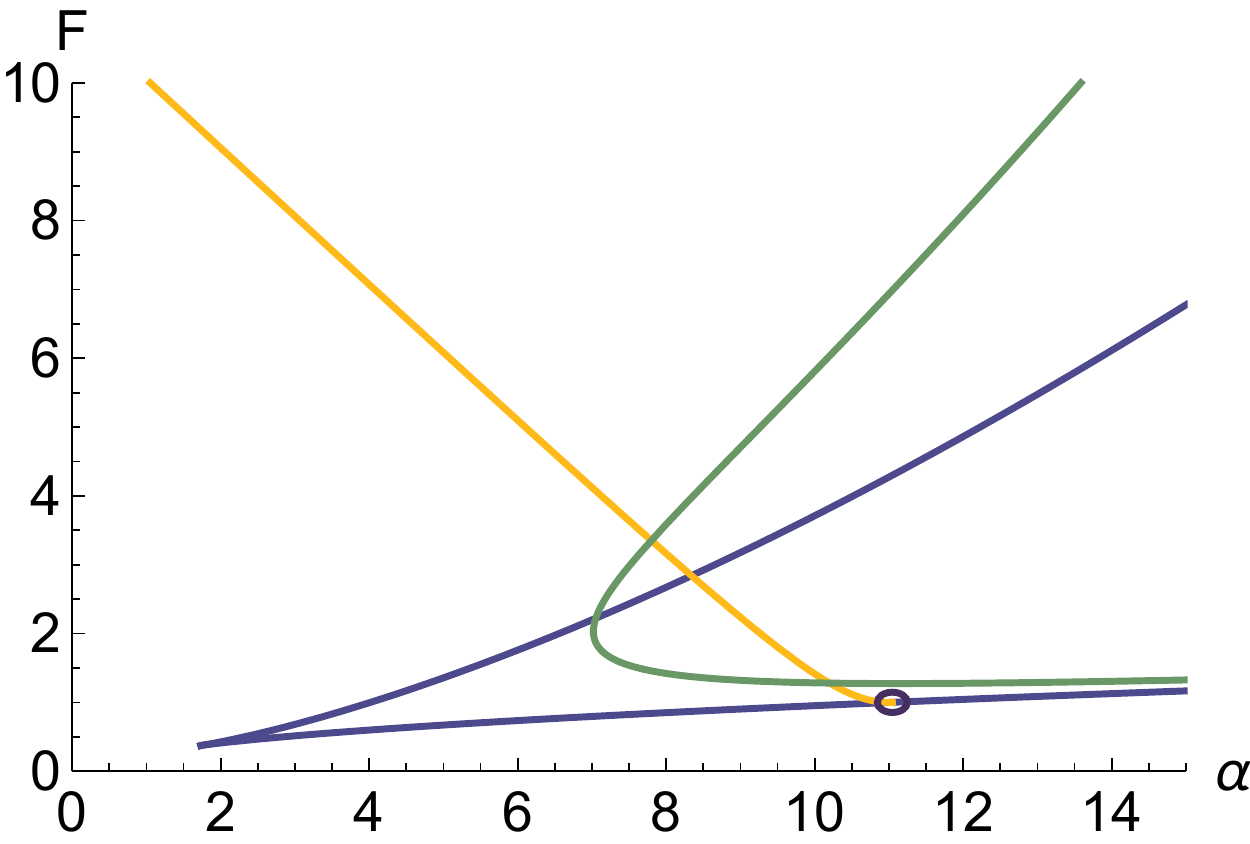}
\caption{\label{fig:alphaf} Stability of continuous wave solutions of the Lugiato-Lefever equation including thermal effects, Eqs.\ \eqref{eq:ll}--\eqref{eq:th}, as a function of pump detuning $\alpha$ and amplitude $F$ for the four sets of thermal parameters shown Fig.\ \ref{fig:alphap}. Each point in the $\alpha$-$F$ plane corresponds to three continuous wave solutions, of which at most two are stable, inside the blue wedge, and to a unique solution (stable or unstable) outside the wedge. The yellow curve marks lower boundary of the region of the finite-wavelength modulational instability, which affects only the upper branch of continuous waves for parameter values in the coexistence wedge in the three panels with $C>0$. The green curve, where relevant, marks the instability threshold of the uniform mode with complex growth eigenvalues, which affects only the upper branch where it overlaps with the coexistence wedge. The dark circles mark the points on the modulational instability boundaries where the fastest growing mode has the largest wavelength. The red dashed line in the upper-right panel shows a deterministic path to access perfect soliton crystals.}
\end{figure}

\begin{figure}[p]
\centering\large\qquad$C=0$\qquad\qquad\qquad\qquad\qquad\qquad$C=5,\,B=0.05$\\
\includegraphics[width=7cm]{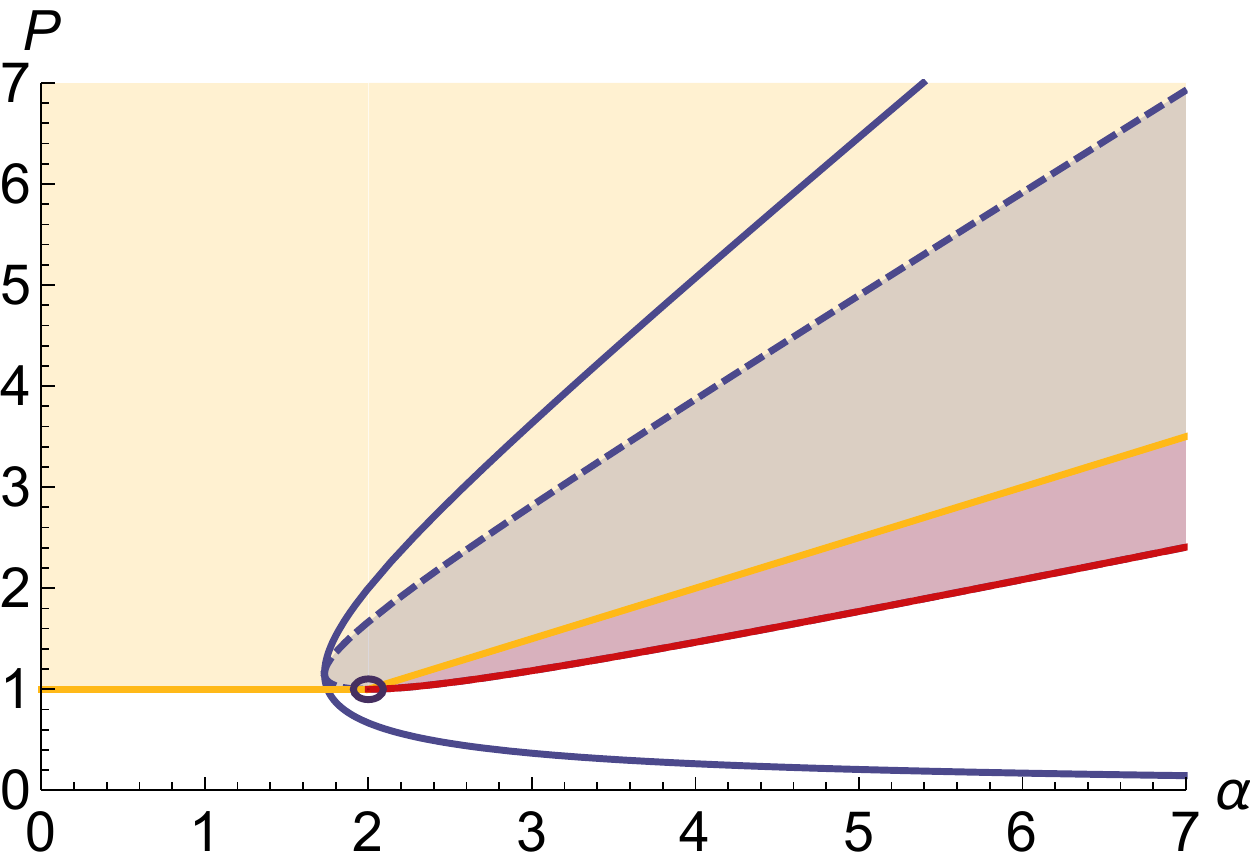}\quad\includegraphics[width=7cm]{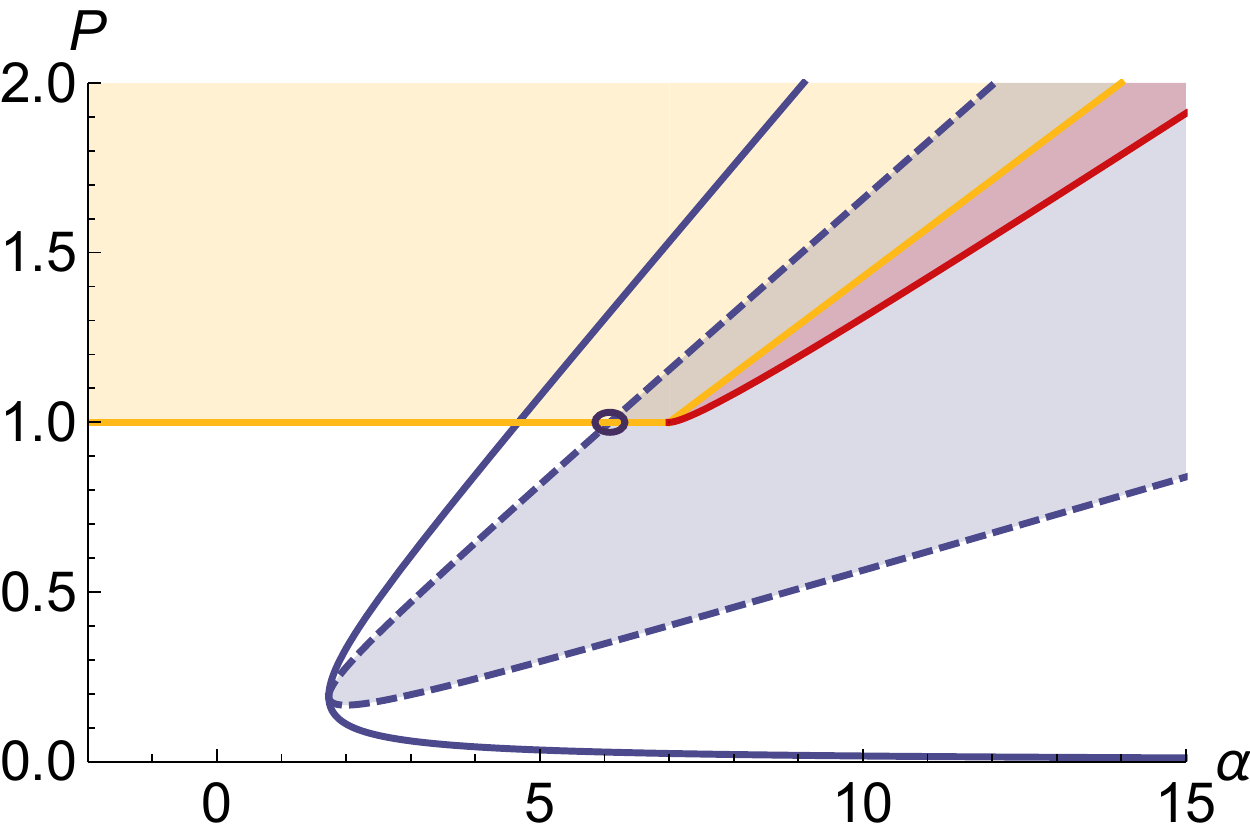}\\[5mm]$C=5,\,B=2$\qquad\qquad\qquad\qquad\qquad$C=10,\,B=0.6$
\includegraphics[width=7cm]{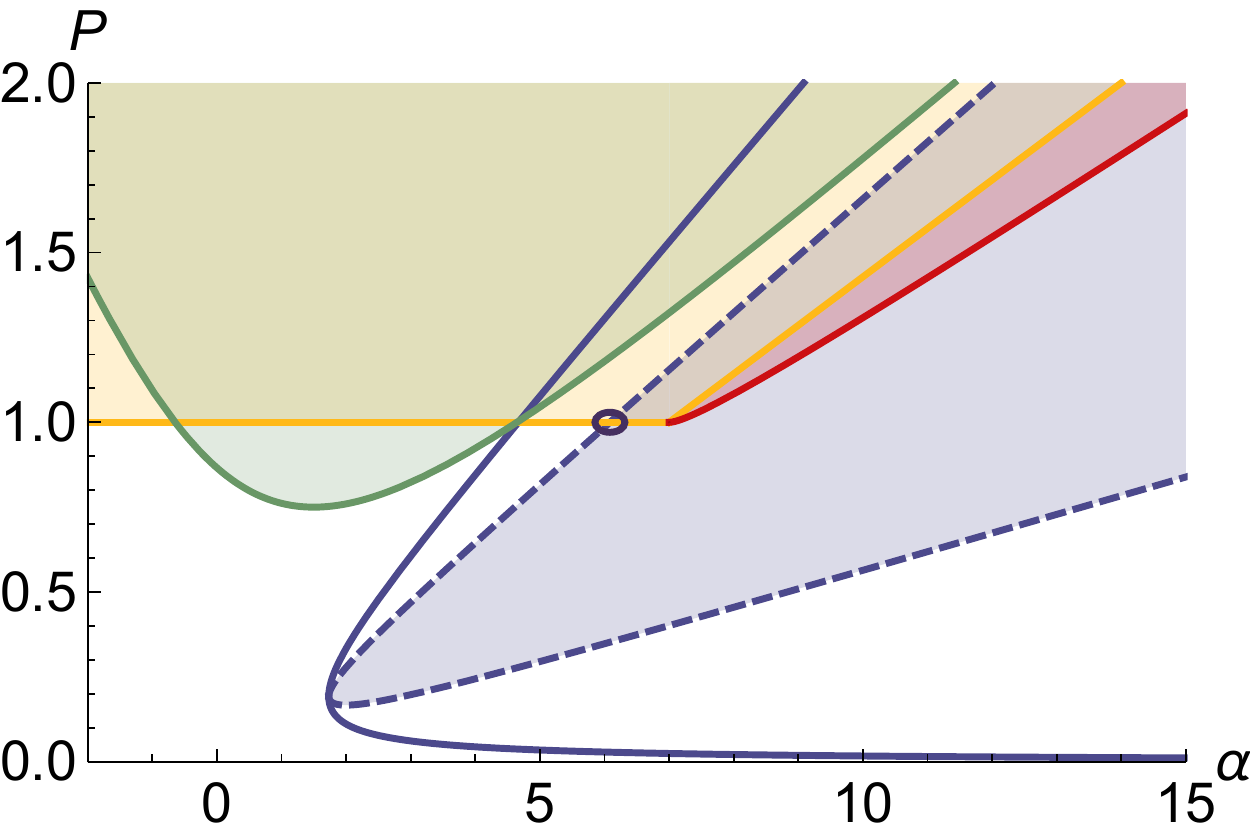}\quad\includegraphics[width=7cm]{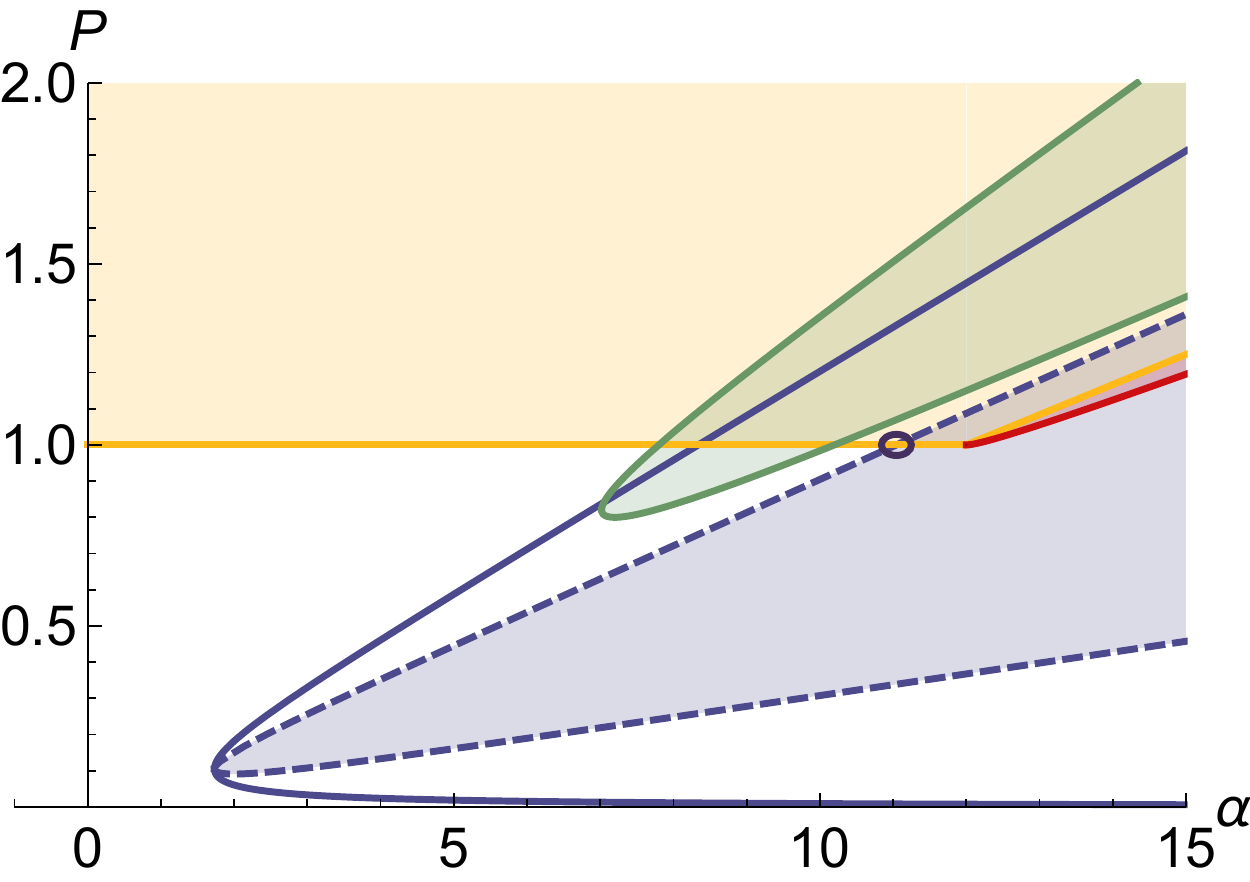}
\caption{\label{fig:alphap} Stability of continuous wave solutions of the Lugiato-Lefever equation including thermal effects, Eqs.\ \eqref{eq:ll}--\eqref{eq:th}, as a function of pump detuning $\alpha$ and mean cavity power $P$ for four sets of thermal parameters. Each point in the $\alpha$-$P$ plane corresponds to a unique continuous wave solution. The solid blue curve is the boundary of the region where three branches exist for the same pump power, and the dashed blue curve marks the boundary between the branches. The solutions in the shaded regions of the graphs are unstable, with shading color indicating the instability modes as follows. Blue: uniform instability with a real growth eigenvalue; red: long-wave modulational instability; yellow: finite-wavelength modulational instability; green: uniform instability with complex growth eigenvalues. The significance of the dark circle is the same as in Fig.\ \ref{fig:alphaf}.
}
\end{figure}

We now distinguish between two cases. When $\alpha<(2+C)P$, we find that $\det M_2$ attains its maximum, $P^2$, at a finite wave number $k_m$; therefore, finite-wavelength instability occurs in the wedge $P>\max\bigl(\alpha/(2+C),1\bigr)$ of the $\alpha$-$P$ plane (shown in yellow in Fig.\ \ref{fig:alphap}). 
The image $F_\text{mi}(\alpha)$ of the line $P=1$ in the $\alpha$-$F$ plane (see Table \ref{tab:defs}) is the square root of a parabola with a minimum of $F=1$ at $\alpha=1+C$ (shown in yellow in Fig.\ \ref{fig:alphaf}). However, when the modulational instability boundary lies inside the coexistence wedge, it is necessary to check the stability properties in the $\alpha$-$P$ plane to determine which of the continuous wave branches it affects. One can verify using the expressions in Table \ref{tab:defs} that for any $C>0$ and for all $\alpha>2+C$, $P_{\text{br}-}(\alpha)<\alpha/(2+C)<P_{\text{br}+}(\alpha)$ , so that the sloping part of the  of the finite-wavelength instability boundary consists of middle-branch points. It follows that for $\alpha\ge2+C$ the modulational instability affects only the upper branch in the coexistence region, considering that the middle branch is inaccessible because it is unstable with respect to uniform perturbations (as discussed Sec.\ \ref{sec:uniform}).

For $\alpha<2+C$, the modulational instability can affect both the lower and upper branches. Nevertheless, the lower branch can become modulationally unstable only if the modulational instability boundary curve intersects the   coexistence region below its tip. This is the behavior for $C=0$ (see Fig.\ \ref{fig:alphap}, top left), but as previously explained, the coexistence region shifts to lower powers as $C$ increases, and in this way the tip of the coexistence region, where the continuous wave branches meet, crosses the modulation instability boundary when $C=C_u=2/\sqrt3-1\approx0.15$. For $C>C_u$, therefore, the modulational instability affects only the upper branch.

This observation, combined with the small magnitude of $C_u$, has an important implication for the deterministic generation of perfect soliton crystals, which utilizes the modulational instability at relatively large red detunings \cite{Qi:2019fla,Kholmyansky:2019cy}. Namely, since red detuned pumps allow for multiple continuous wave branches, and having observed that for typical thermal conditions the lower branch is not susceptible to the modulational instability, the deterministic access path must begin at a relatively small detuning in order to bypass the tip of the coexistence region, before the pump detuning is increased toward the modulational instability boundary. The maximal detuning that can be achieved in this way, $\alpha_t$ (expression given in Table \ref{tab:defs}), is where the modulational stability threshold curve crosses the boundary between the upper and middle branches in the $\alpha$-$P$ plane; this point, which corresponds to the tangency between modulational instability curve and the lower boundary of the coexistence wedge in the $\alpha$-$F$ plane, is circled in both Figs.\ \ref{fig:alphap} and  \ref{fig:alphaf}. The deterministic path to access \hl{perfect soliton crystals} is discussed further in Sec. \ref{sec:cnoidal}.

The second case of the modulational instability occurs when $\alpha\ge(2+c)P$. The maximum of $\det M_2$ is then at $k=0$, so that the fastest growing mode of this instability has the smallest admissible positive wave number, $k_{\min}=2\pi/L$. In applications, the resonant mode length $L$ is typically larger than 10 in system units \cite{Qi:2019fla}, so that $\max\det M_2\approx\det M_2(k=0)$. It follows that the long-wave instability affects continuous waves with $\det M_2(k=0)>1$ and $\alpha\ge(2+C)P$. The solution of the quadratic inequality $\det M_2(k=0)>1$ is $P_{\text{lw}-}<P<P_{\text{lw}+}$, (see  Table \ref{tab:defs} for definitions). 

It follows from \eqref{eq:detm2} that $P>1$ is a necessary condition for instability, and therefore $\alpha>2+C$ is a necessary condition for the long-wave instability. On the other hand, it can be checked using the expressions in Table \ref{tab:defs} that $P_{\text{lw}-}<\alpha/(2+C)<P_{\text{lw}+}$ for $\alpha>2+C$, so that the region in $\alpha,\,P$ parameter space for long-wave instability is bounded from above by $\alpha/(2+C)$ and from below $P_{\text{lw}-}$ (defined in Table \ref{tab:defs}); the lower boundary curve is marked in red in Fig.\ \ref{fig:alphap}. It also follows from the expressions in Table \ref{tab:defs} that $P_{\text{br}-}<P_{\text{lw}-}$ and $P_{\text{lw}+}<P_{\text{br}+}$ for any $C>0$, so that the long-wave instability only affects the physically inaccessible middle branch.

\subsection{Uniform instabilities}\label{sec:uniform}
We now consider the case where $\psi_1(t)$ is spatially uniform. In this case \hl{$\Theta_1$ is nonzero, while} the dispersion term in Eq.\ \eqref{eq:ll1} drops, and we obtain the \hl{three-variable} system
\begin{align}\label{eq:mat}
&\frac{d}{dt}\begin{pmatrix}\psi_1\\\psi_1^*\\\Theta_1\end{pmatrix}=M_3\begin{pmatrix}\psi_1\\\psi_1^*\\\Theta_1\end{pmatrix}\\&M_3=
\begin{pmatrix}-1+i(-\alpha+(2+C)|\psi_c|^2)&i\psi_c^2&-i\psi_c\\-i(\psi_c^*)^2&-1-i(-\alpha+(2+C)|\psi_c|^2)&i\psi_c^*\\-A\psi_c^*&-A\psi_c&-B\end{pmatrix}\label{eq:m3}\end{align}
The eigenvalues of $M_3$  are the roots of a cubic polynomial, so that there are two cases: \hl{$M_3$ has either} three real eigenvalues, or one real eigenvalue and a complex conjugate pair. In the former case stable solutions are characterized by three negative eigenvalues, and in the latter case by one negative eigenvalue, and a pair of complex conjugate eigenvalues with negative real parts, so that $\det M_3<0$ for all stable solutions.

A real-eigenvalue instability occurs when the largest of the real eigenvalues, and therefore $\det M_3$, changes sign. A direct calculation starting from \eqref{eq:m3} shows that $\det M_3>0$ whenever $P_{\text{br}-}<P<P_{\text{br}-}$, that is, precisely for the middle branch, which corresponds to the interior of the dashed-blue lobe in Fig.\ \ref{fig:alphap}. The entire middle branch is therefore unstable; this instability, which generalizes an analogous instability in nonthermal resonators \cite{Godey:2014cj}, is typical for the middle branch in bistable systems \cite{Strogatz:1994tz}. The middle-branch solutions are not close to any stable solutions except where the branch connects to the lower and upper branches in a standard saddle-node bifurcation \cite{Strogatz:1994tz}. 

A complex-eigenvalue instability occurs when the real part of the complex conjugate pair of eigenvalues changes sign. It is a finite-frequency instability with an oscillatory growth mode that is created in a Hopf bifurcation. We show below in Sec. \ref{sec:osc} that the finite-frequency instability growth saturates nonlinearly, yielding stable periodic temporal oscillations of spatially uniform waves. 

For finite-frequency unstable solutions, the matrix $M_3$ has two complex conjugate eigenvalues with positive real parts and one negative eigenvalue. Such a combination of eigenvalues can only occur when $A>2$, and then there are two cases, depending on whether $A$ is smaller or larger than $3+C$. In the former case, a finite-frequency instability occurs for detunings $\alpha>\alpha_h$, and powers in the interval $P_{h-}<P<P_{h+}$; $\alpha_h$ and $P_{h\pm}$ are defined in Table \ref{tab:defs}. We show an example in the bottom-right panels of Figs.\ \ref{fig:alphaf}, \ref{fig:alphap}. In the latter case the instability can occur for any $\alpha$ for $P>P_{h-}$;   we show an example in the bottom-left panels of Figs.\ \ref{fig:alphaf}, \ref{fig:alphap}.

\section{Cnoidal waves}\label{sec:cnoidal}
The growth of finite-wavelength perturbations in modulationally unstable continuous waves can saturate to yield stable stationary spatially periodic waves, also known as cnoidal waves, Turning rolls, and perfect soliton crystals. In practice, cnoidal waves can be accessed by fixing the pump detuning $\alpha$ and raising its amplitude $F$ to the point where continuous waves become modulationally unstable.

When thermal effects are negligible ($C=0$) there are two cases: If $\alpha<\alpha_c(0)=41/30$, the bifurcation is supercritical, which means that when $F$ is slightly above the instability threshold $F_\text{mi}$, the growth saturates to a cnoidal wave that is a quasiharmonic perturbation of the continuous wave solution,
\begin{equation}\label{eq:scn}
\psi_s(x)=\psi_c+\psi_a\cos(k_m x+\phi)\ ,
\end{equation}
where $k_m$ is the wave number of the fastest growing mode, $\phi$ is an arbitrary phase, and the amplitude \cite{ParraRivas:2018ie}
\begin{equation}
\psi_a\propto\sqrt{F-F_\text{mi}}\ .
\end{equation}
When $\alpha>\alpha_c$, there are no stationary stable waveforms that closely approximate continuous waves. 
Nevertheless instability growth can saturate in cnoidal waves whose  amplitude variations are not small \cite{Kholmyansky:2019cy}.

\begin{figure}
\includegraphics[width=8cm]{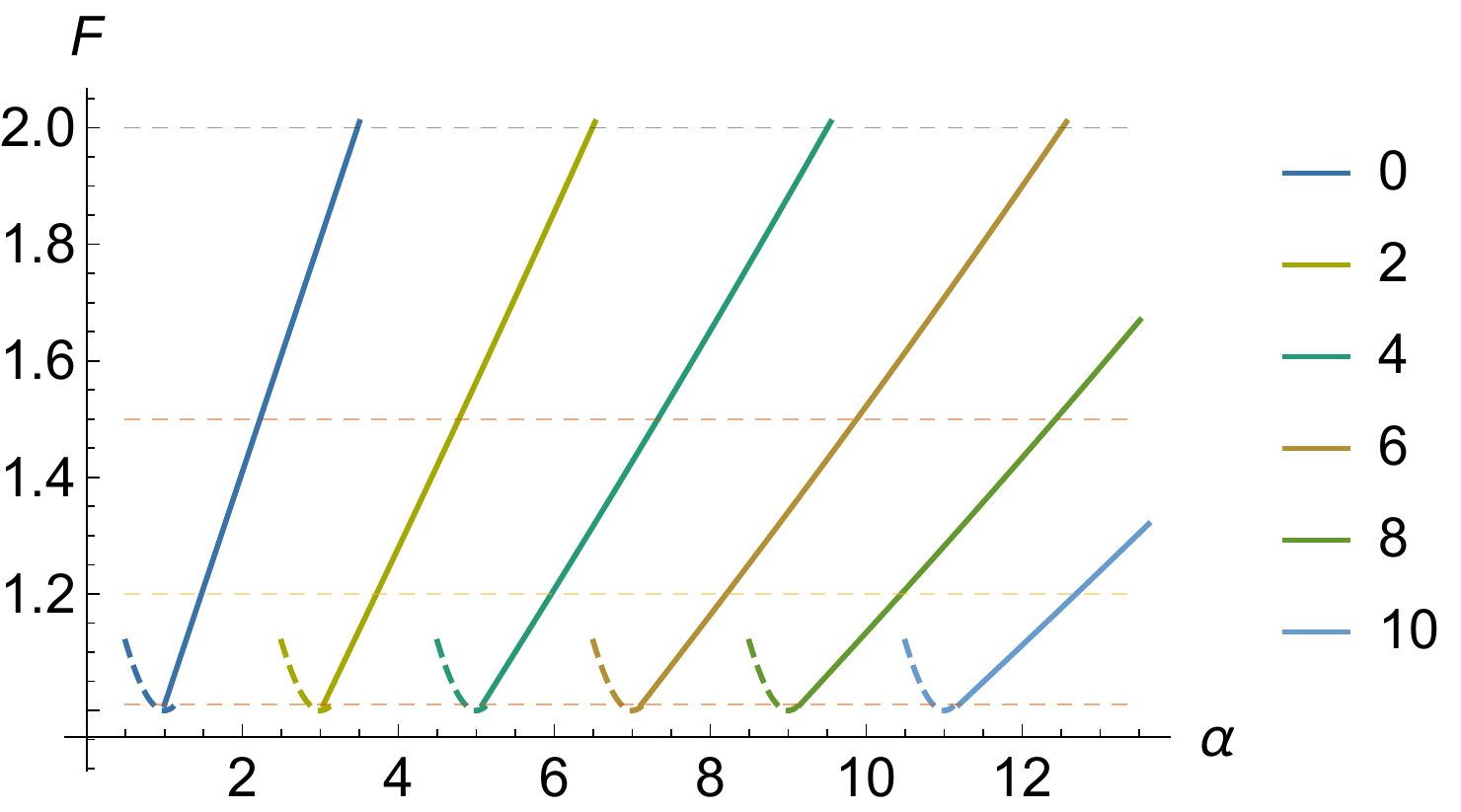}\quad\includegraphics[width=8cm]{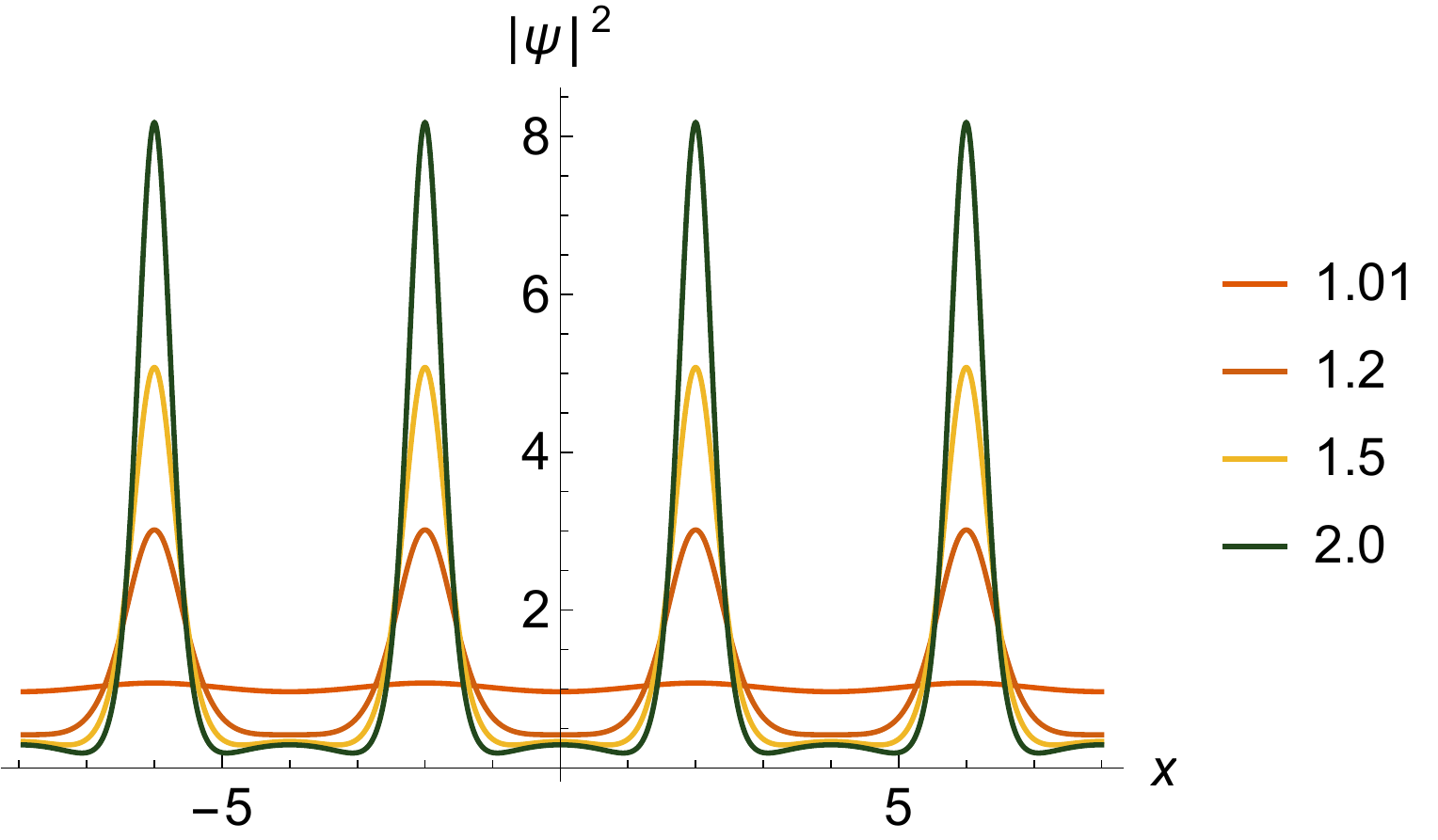}\caption{\label{fig:cnw}
Left: Full lines show deterministic access paths in the pump parameter space leading from the lower tip of the modulational instability curve to soliton crystals for several choices of the thermal parameters $A=C=0,\,2,\,4,\,6,\,8,\,10$ from left to right. The access paths are related by the exact mapping \eqref{eq:smap}, so that cnoidal waveforms with the same $F$ are identical. Stability properties, however, depend on the thermal parameters, and the cnoidal waves on the access paths for $C=8$ and $C=10$ become unstable before reaching $F=2$. The horizontal dashed lines show the values of $F$ for which the intensity profiles are shown in the right panel. Dashed curves show a small part of the modulational instability threshold for the respective values of $C$. Right: The intensity profile $|\psi_s(x)|^2$ for stable cnoidal waves at several points on the access paths shown in the left panel, located at $F=1.01,\,1.2,\,1.5,\,2.0$ (where applicable). }
\end{figure}

Like all stationary waveforms, a cnoidal wave with $C>0$  can be mapped via \eqref{eq:smap} to a non-thermal cnoidal wave with a smaller detuning $\alpha_0=\alpha-CP_s$, so the set of possible cnoidal waves (but not their stability) is independent of $C$.
In general, since $P_s$ depends in a complicated manner on $\alpha$ and $F$,  the mapping \eqref{eq:smap} cannot easily be used to relate the stationary cnoidal wave for different thermal parameters.  
Nevertheless, weakly nonlinear cnoidal waves near the supercritical part of the modulational instability curve have mean power $P_s\approx P_c\approx 1$ as discussed after Eq.\ \eqref{eq:detm2}, and the threshold curves (yellow lines in Fig.\ \ref{fig:alphaf}) are shifted to larger $\alpha$ as $C$ increases. It now follows from \eqref{eq:smap} that this rule remains true also in a small region \emph{above} the threshold. That is, if a cnoidal wave \eqref{eq:scn} is a stationary solution for some $\alpha_0$ and $F$ close to $F_\text{mi}$ at $C=0$, then it is also a stationary solution for the same $F$ and $\alpha=\alpha_0+C$ for any $C\ge0$. In particular it follows that the bifurcation is supercritical for a given $C$ when
\begin{equation}\label{eq:alphac}
\alpha<\alpha_c(C)=\frac{41}{30}+C\ .
\end{equation}

This simple result has a significant implication: As explained above, broadband frequency combs with soliton-like envelopes can be produced from highly red-detuned perfect soliton crystals, and can be deterministically accessed from long-wave modulational instabilities \cite{Qi:2019fla,Kholmyansky:2019cy}. When $C=0$ the largest wavelength that can be accessed in this way is produced by crossing the instability curve at $\alpha=\alpha_c$. However, as $C$ increases the detuning $\alpha_t$ at the tangency point between the modulational instability curve and the cw coexistence wedge (circled in Fig.\ \ref{fig:alphaf}, see Sec.\ \ref{sec:mod} above and Table \ref{tab:defs}) increases more slowly than $\alpha_c$, so that $\alpha_c$ overtakes $\alpha_t$ when $C=C_s=361/660\approx0.55$. For $C>C_s$ therefore, the modulational instability curve consists entirely of supercritical bifurcation points, and the optimal deterministic path in the $\alpha$-$F$ parameter space for accessing soliton-like cnoidal waves is via continuous wave solutions with detuning $\alpha_t$.

Since $C_s>C_u$ (see Sec.\ \ref{sec:mod}), the modulational instability for $C>C_s$ affects the \emph{upper} branch of continuous wave solutions. It follows that  for the deterministic generation of wideband frequency combs in any microresonator with an appreciable thermal response, the pump power has to be raised at a fixed frequency with $\alpha<\sqrt3$, in order to bypass the wedge of coexistence in the pump parameter space. Then the pump frequency has to be sharply detuned to the red at fixed power, to reach the optimal access point with $\alpha=\alpha_t\approx1+C$. We show an example in the top-right panel of Fig. \ref{fig:alphaf}.

After crossing the modulational instability threshold and the consequent formation of a quasiharmonic cnoidal wave, soliton crystals can be obtained by further increasing the power and detuning of the pump \cite{Qi:2019fla,Kholmyansky:2019cy}. Figure \ref{fig:cnw} (left) shows several access paths for different values of $C$, chosen to give identical waveforms that are connected by the exact mapping \eqref{eq:smap} for equal pump amplitudes. The right panel of Fig.\ \ref{fig:cnw} shows cnoidal wave solutions at several points along the access paths. 

\begin{figure}[p]\large$\qquad\qquad\qquad\qquad B=2,\,C=5,\,\alpha=8 \qquad\qquad\! B=0.6,\,C=10,\,\alpha=15$\\[5mm]
\large \raisebox{2cm}{$F_\Delta=0.0005$}\ \includegraphics[width=6.7cm]{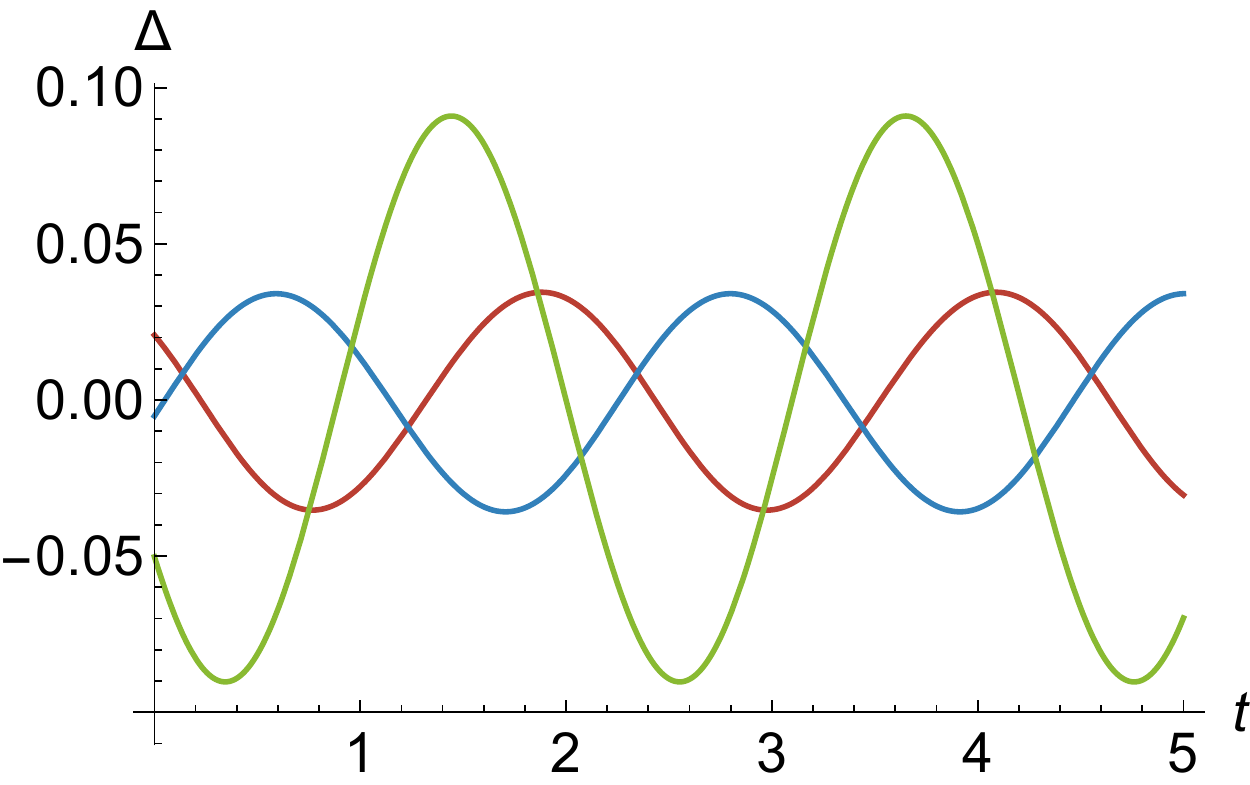}\ \includegraphics[width=6.7cm]{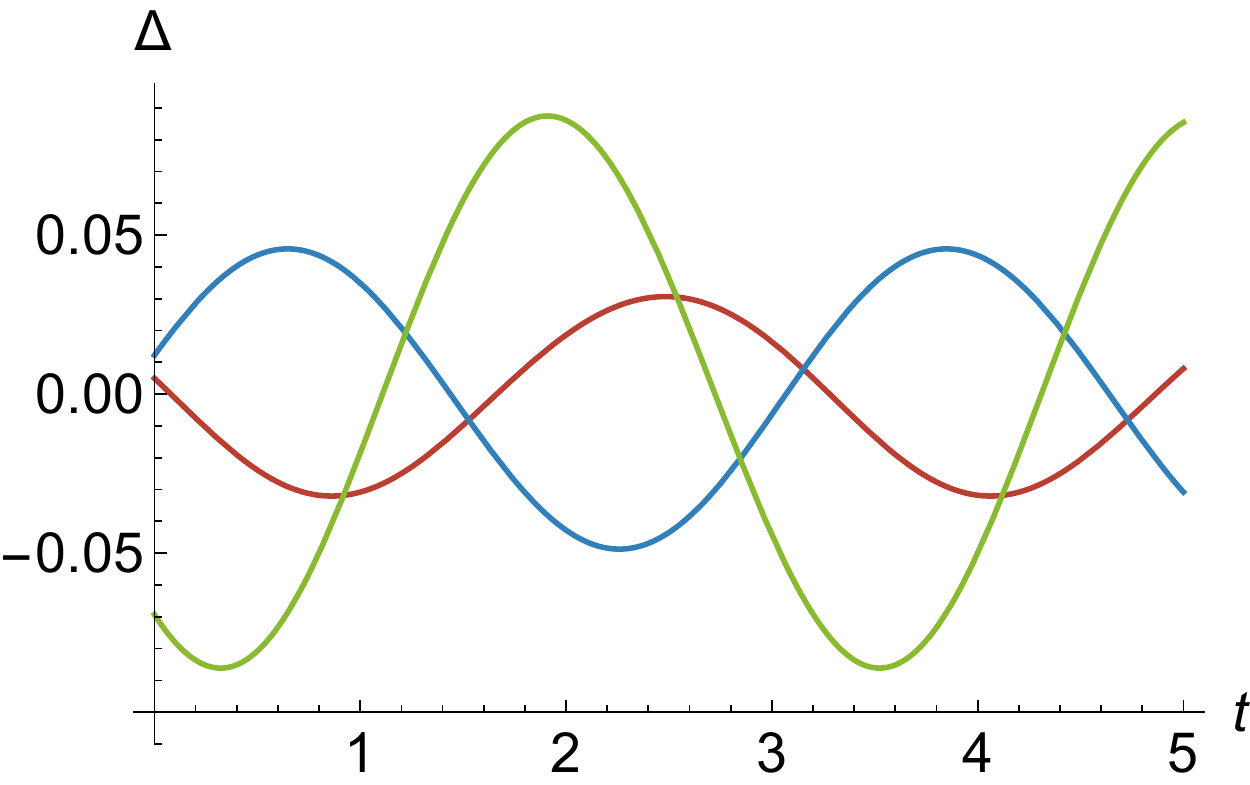}\\[5mm]
\large \raisebox{2cm}{$F_\Delta=0.1$}\qquad\includegraphics[width=6.7cm]{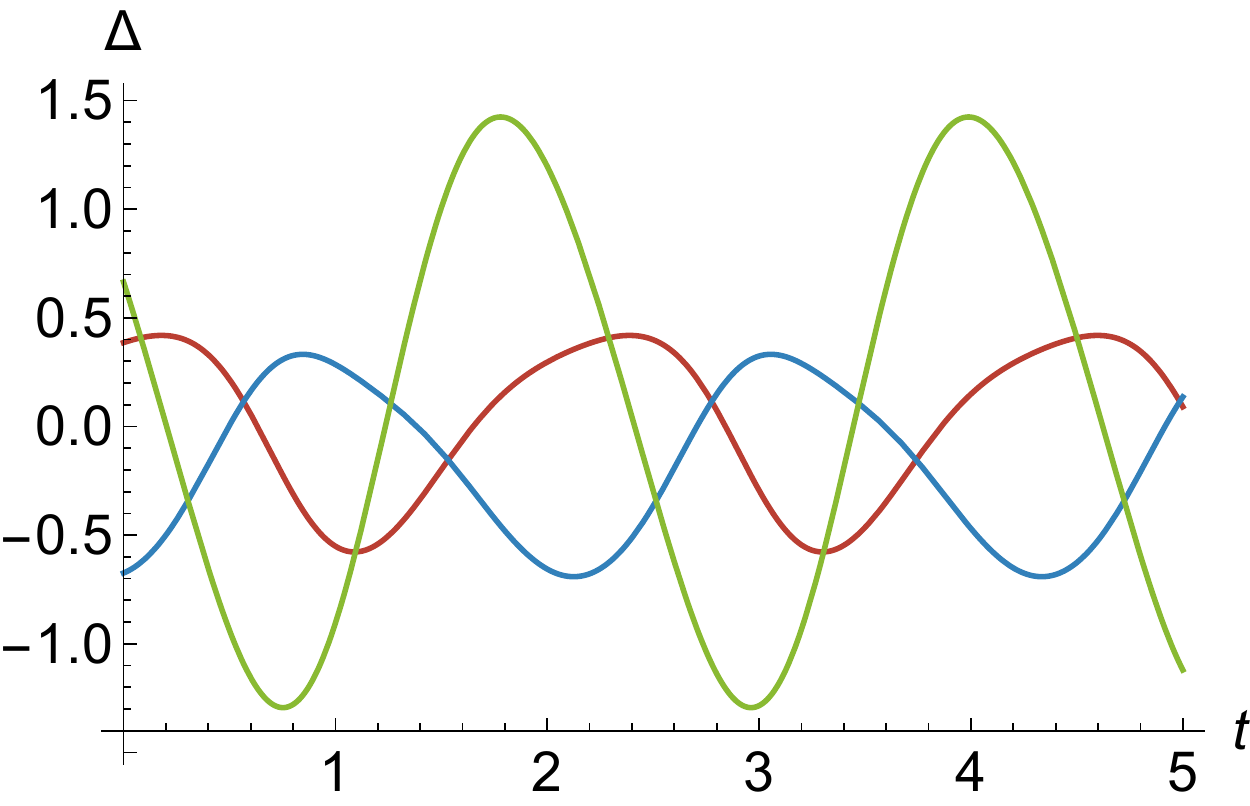}\ \includegraphics[width=6.7cm]{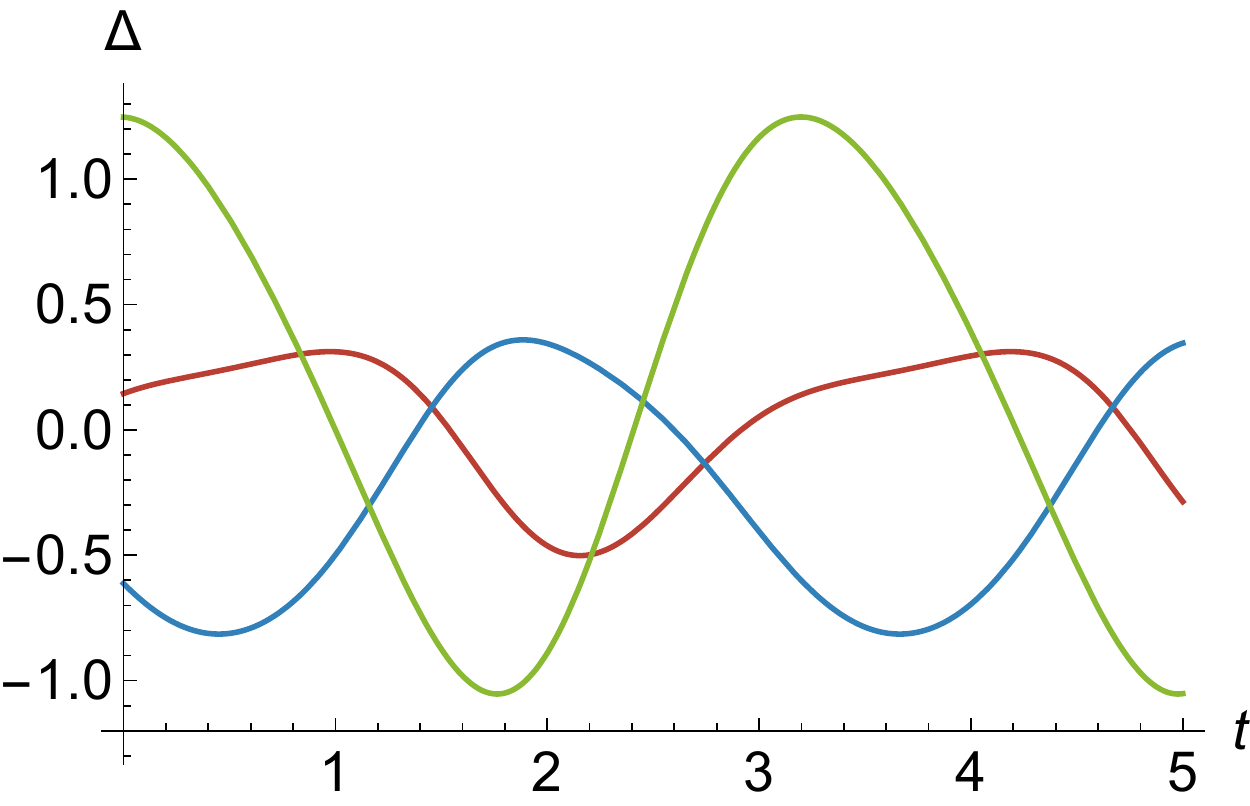}\\[5mm]
\large \raisebox{2cm}{$F_\Delta=2.0$}\qquad\includegraphics[width=6.7cm]{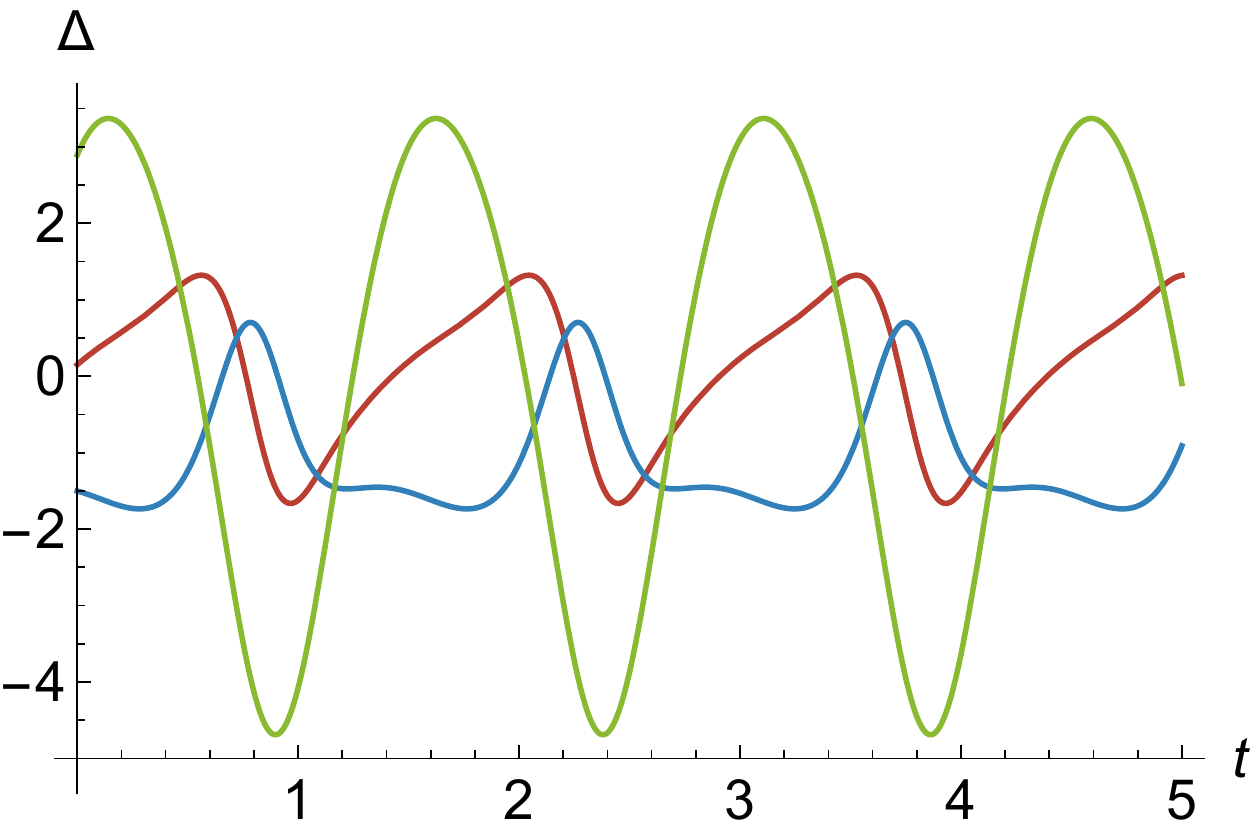}\;\;\includegraphics[width=6.7cm]{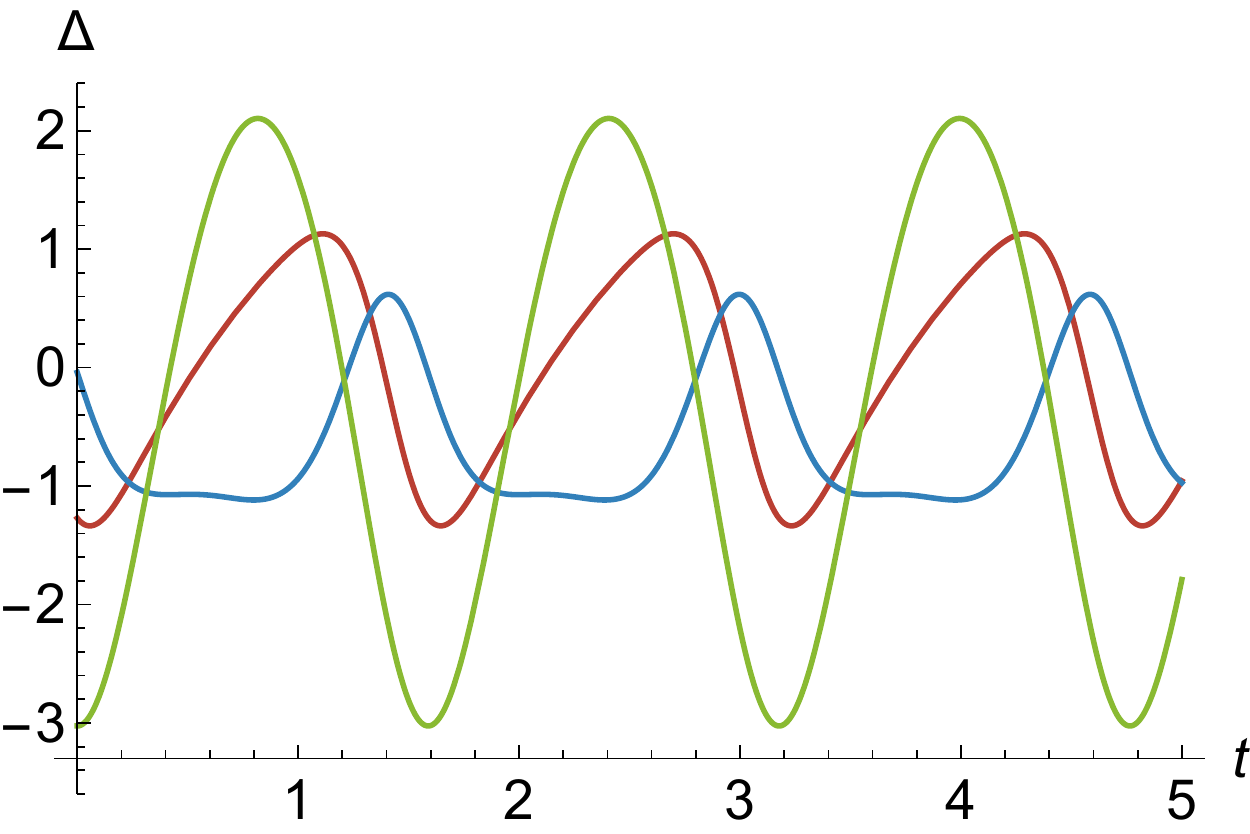}
\caption{\label{fig:osc}Steady state thermal oscillations of a spatially uniform wave envelope field in a Kerr resonator. The deviations $\Delta$ of the field envelope $\psi-\psi_c$ (real part in red, imaginary part in blue) and of the thermal detuning $\Theta-\Theta_c$ (green) from the constant values are shown as a function of time $t$. The left column shows oscillation with thermal parameters $B=2,\,C=5$ and detuning $\alpha=8$, and the right column shows oscillation with $B=0.6,\,C=10$ and $\alpha=15$. The pump amplitude is $F=F_h+F_\Delta$, where $F_h$ is the instability threshold, and $F_\Delta=0.0005,\,0.1,\,\text{and }2.0$ in the top, middle, and bottom rows, respectively. The quasi-harmonic oscillations of the top row are well-described by weakly nonlinear analysis. (Note different vertical axes scales.)}
\end{figure}

\section{Thermal oscillations}\label{sec:osc}
In Sec. \ref{sec:uniform} we found that when $A>2$, there is a region in the pump-parameter plane where continuous waves are unstable with a complex conjugate pair of eigenvalues with positive real parts. The boundary of the region of instability in the $\alpha$-$F$ parameter space is a line of Hopf bifurcations that occur when a pair of stability eigenvalues $\pm i\omega_h$ cross the imaginary axis. It follows from the Hopf bifurcation theorem \cite{Strogatz:1994tz} that a periodic solution of the equations of motion with frequency $\omega_h$ is created at the bifurcation points.

There are two types of Hopf bifurcations: When the bifurcation is supercritical, a stable periodic orbit is created on the unstable side, while at a subcritical bifurcation an unstable periodic orbit is created on the stable side.
The oscillatory instability occurs when the cw power $P_c$ becomes larger than the lower threshold $P_{h-}$. Close to the bifurcation the oscillations are quasiharmonic with amplitude proportional to $\sqrt{|P_\Delta|}$, where $P_\Delta=P_c-P_{h-}$. 

The quasiharmonic oscillations are well-approximated by the linear modes of the stability matrix $M_3$, with an amplitude that is determined by the balance of the linear and nonlinear terms. Since the amplitude is small, the nonlinearity is weak, and therefore the solution can be calculated in weakly nonlinear perturbation theory \cite{Strogatz:1994tz}. The result of this calculation is that close to the bifurcation
\begin{equation}
\begin{pmatrix}\psi(t)\\\psi(t)^*\\\Theta(t)\end{pmatrix}=\begin{pmatrix}\psi_c\\\psi_c^*\\-C|\psi_c|^2\end{pmatrix}+\sqrt{\pm P_\Delta}\bigl(Kv_+e^{i\omega t}+K^*v_-e^{i\omega t})\ ,
\end{equation}
where $\psi_c$ is the continuous wave amplitude, $K$ is a constant, $\omega=\omega_h+O(P_\Delta)$, and $v_\pm$ are the $\pm i\omega_h$ normalized eigenvectors of the matrix $M_3$ (see Eq.\ \ref{eq:m3}) with phase chosen to make their $\Theta$ component real. The sign under the square root is positive (resp.\ negative) for supercritical (resp.\ subcritical) bifurcations, $|K|$ is fixed by the nonlinear terms, while the phase of $K$ is determined by the initial conditions.

\begin{figure}[t]{\hspace{29mm}\large$C=5$\hspace{62mm}\large$C=10$\hspace{30mm}\large$|K|$}\\
\hbox{\includegraphics[width=7.cm]{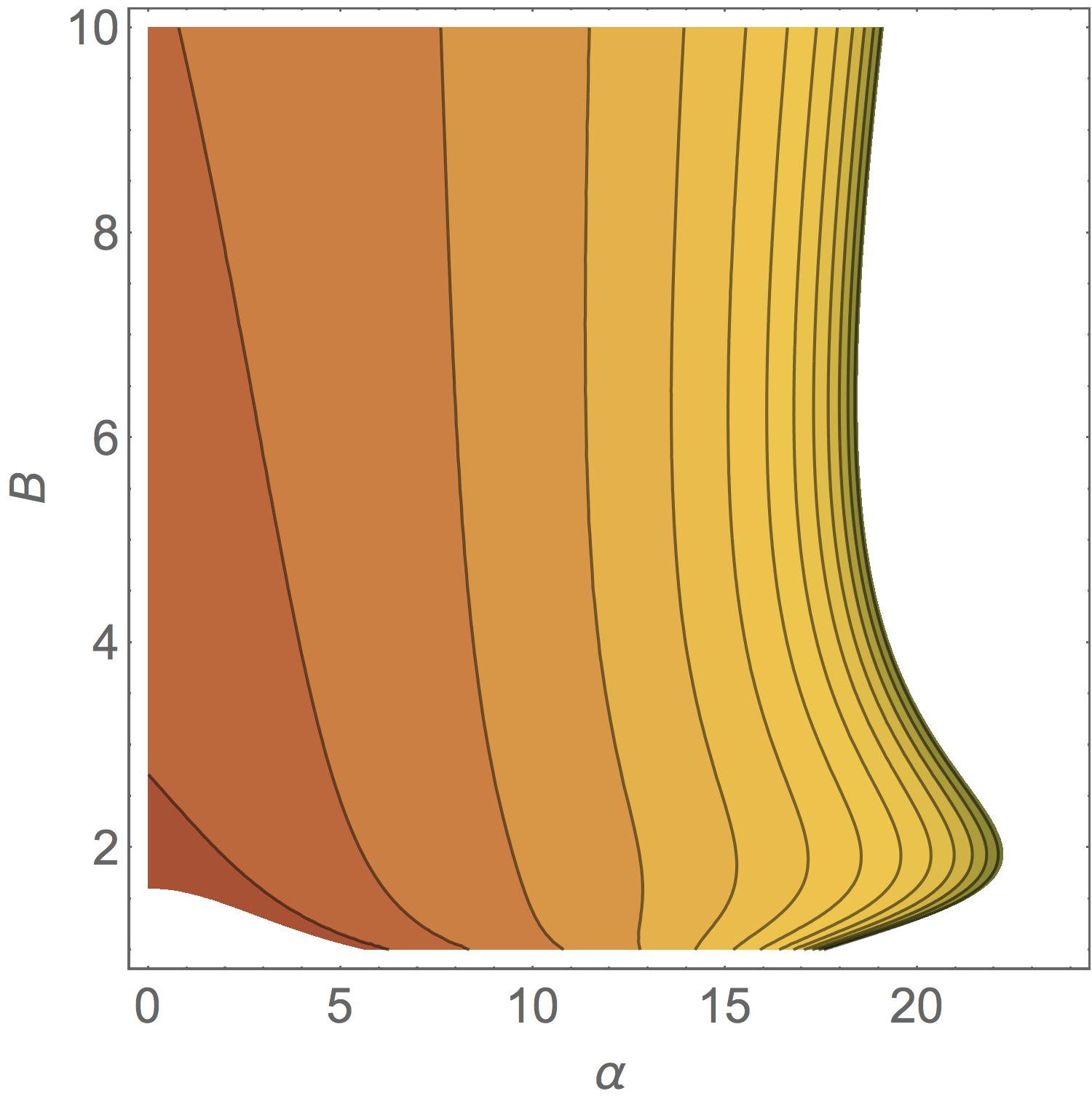}\quad\includegraphics[width=7.cm]{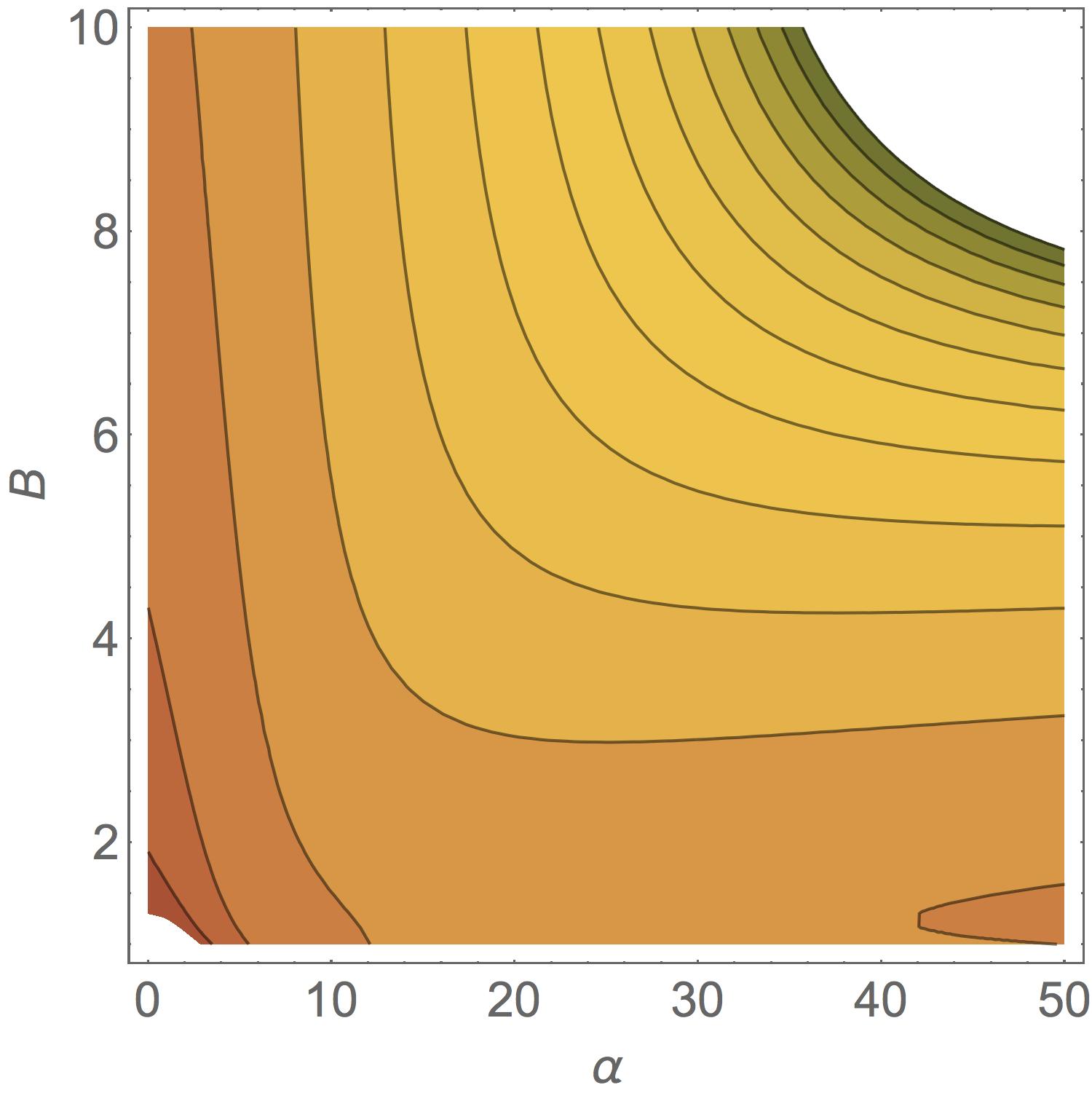}\quad\includegraphics[height=7.3cm]{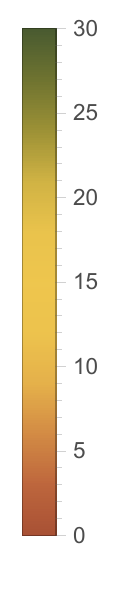}}
\caption{\label{fig:wnl}The nonlinear response parameter $|K|$ of the uniform instability Hopf bifurcation as a function of the pump detuning $\alpha$ and cooling rate $B$ for thermal sensitivity ratio $C=5$ (left) and $C=10$ (right). The plots are colored in parameter regions where the bifurcation is supercritical. Uncolored regions correspond either to points where the bifurcation is subcritical (large $\alpha$ regions) or parameter values where the Hopf bifurcation does not occur at all (small $\alpha$ regions.)  }
\end{figure}

In Fig.\ \ref{fig:osc} we show examples of stable oscillatory solutions to the dynamical equations \eqref{eq:ll}--\eqref{eq:th}. The panels of the first row show quasi-harmonic oscillations for pump amplitudes slightly above threshold. When the pump power increases further, the oscillations quickly become strongly nonlinear, as seen in the second and third rows of Fig.\ \ref{fig:osc}.
In Fig.\ \ref{fig:wnl} we show the nonlinear response coefficient $|K|$ of quasiharmonic oscillations as a function of the pump detuning $\alpha$ and cooling coefficient $B$ for two values of $C$. 

The weakly nonlinear analysis reveals that there are parameter values where the Hopf bifurcation is subcritical, for example in the white regions in the right side of both panels of Fig.\ \ref{fig:wnl}. For these parameter values there are no small-amplitude stable thermal oscillations near the threshold. 
Nevertheless, in a number of these cases where we studied  numerically the growth of instability following a subcritical bifurcation, we always found the it saturates at strongly nonlinear stable oscillations. These thermal oscillations are qualitatively similar to the  oscillations shown in the second and third rows of Fig.\ \ref{fig:osc}.

\section{Conclusions}\label{sec:conclusions}
Dissipative heating plays a significant role in the evolution of optical waveforms in Kerr microresonators. 
For theoretical modeling of this effect, the Lugiato-Lefever equation must be coupled to an equation for the mean temperature of the optical mode volume. 
The temperature evolution then depends on two response coefficients, describing the rate of heating due to absorption of light, and the rate of cooling due to diffusion. The ratio of these coefficients is the dimensionless thermal sensitivity parameter $C$.
When $C$ is of order one or larger, as it is for most Kerr microresonators in which frequency combs have been observed, the existence and stability properties of cavity waveforms depend strongly on the thermal parameters.

\hl{Here we focused on continuous waves and cnoidal waves, where for pump parameters $\alpha$ and $F$ of order one, the mean power $P$ is also of order one. Single soliton waveforms on the other hand have mean power $P=1/L$, so that the strength of the thermal effects for single solitons is determined by $C/L$ which is typically small. It means however, that thermal effects for single soliton waveforms can become significant when $L$ decreases, which can happen either when the cavity is smaller, or when the dispersion is larger.}

An important simplifying feature of the thermooptics of microresonators is that steady-state waveforms experience an effective detuning that is shifted from the nominal detuning by $C$ times the mean power, so that there is an exact mapping between the set of thermal and non-thermal steady-state waveforms. However, the \emph{stability} properties of the waveforms depend on the details of the thermal response, and cannot in general be deduced from those of the nonthermal system.

We showed that dissipative heating has three important consequences for the stability of continuous waves and the evolution of the cavity waveform following the onset of instability. First, as the thermal sensitivity parameter increases, the coexistence wedge, where there are three possible solutions, moves to lower powers in the pump parameter plane, blocking the constant-frequency path to red-detuned comb-producing waveforms. At the same time, the modulational instability threshold curve, where comb-forming cnoidal waves bifurcate from continuous waves, moves to higher detuning. Together, these two effects imply that soliton crystals can be produced {deterministically} by adiabatically following a path of pump parameters that  bypasses the coexistence wedge by first increasing pump power at constant frequency and then tuning the frequency to the red at constant power. This protocol has not yet been implemented experimentally.

The third consequence of the thermal response is the emergence of an instability mode of the continuous waves that produces temporally periodic oscillations of spatially uniform waveforms. Thermal oscillations in Kerr resonators have been often observed experimentally, and our analysis shows how to describe them with a simple three-degrees of freedom dynamical system. We showed that oscillations can only occur when the heating rate is larger than a threshold value which is universal in dimensionless units. \hl{It is likely that the mechanism that drives the thermal instabilities and oscillations in continuous waves can cause similar instabilities and oscillations of cnoidal waves, and perhaps soliton waveforms as well, but this question is beyond the present scope.}

In experiments, an effective heating rate large enough to produce thermal oscillations has been achieved by lowering the response coefficient $d\omega_c/dT$ which appears in its denominator (see Eq. \ref{eq:thpar}), either by a cancellation between competing effects as in \cite{Diallo:2015ck} or by  operating near the temperature where the response coefficient changes sign \cite{Park:2007wf}. Thermal oscillations in experiments that have been reported so far were strongly nonlinear. The analysis presented here shows that for a wide range of parameters the Hopf bifurcation that gives rise to the oscillations is supercritical, so that weakly nonlinear quasiharmonic oscillations are expected as well; however, the parameter band where the oscillations are quasiharmonic is narrow so that a careful and slow tuning of the pump parameters through the bifurcation will be needed to observe them experimentally. 

\hl{\paragraph{Acknowledgments}
AL and OG thank the US-Israel Binational Science Foundation for financial support under NSF-BSF grant number 2017643. ZQ, TFC, and CRM thank the National Science Foundation for financial support under grant ECCS-1807272.}


\begin{thebibliography}{10}

\bibitem{Pfeiffer:2017ia}
M.~H.~P. Pfeiffer, C.~Herkommer, J.~Liu, H.~Guo, M.~Karpov, E.~Lucas,
  M.~Zervas, and T.~J. Kippenberg, ``{Octave-spanning dissipative Kerr soliton
  frequency combs in Si$_3$N$_4$ microresonators},'' {\em Optica}, vol.~4,
  no.~7, p.~684, 2017.

\bibitem{Spencer:2018eb}
D.~T. Spencer, T.~Drake, T.~C. Briles, J.~Stone, L.~C. Sinclair, C.~Fredrick,
  Q.~Li, D.~Westly, B.~R. Ilic, A.~Bluestone, N.~Volet, T.~Komljenovic,
  L.~Chang, S.~H. Lee, D.~Y. Oh, M.-G. Suh, K.~Y. Yang, M.~H.~P. Pfeiffer,
  T.~J. Kippenberg, E.~Norberg, L.~Theogarajan, K.~Vahala, N.~R. Newbury,
  K.~Srinivasan, J.~E. Bowers, S.~A. Diddams, and S.~B. Papp, ``{An
  optical-frequency synthesizer using integrated photonics},'' {\em Nature},
  vol.~557, pp.~81--85, Apr. 2018.

\bibitem{Kippenberg:2018hi}
T.~J. Kippenberg, A.~L. Gaeta, M.~Lipson, and M.~L. Gorodetsky, ``{Dissipative
  Kerr solitons in optical microresonators},'' {\em Science}, vol.~361,
  p.~eaan8083, Aug. 2018.

\bibitem{Chembo:2015hz}
Y.~K. Chembo, ``Kerr optical frequency combs: theory, applications and
  perspectives,'' {\em Nanophotonics}, vol.~5, pp.~214--230, June 2016.
\newblock Publisher: De Gruyter Section: Nanophotonics.

\bibitem{Pasquazi:2018de}
A.~Pasquazi, M.~Peccianti, L.~Razzari, D.~J. Moss, S.~Coen, M.~Erkintalo, Y.~K.
  Chembo, T.~Hansson, S.~Wabnitz, P.~Del’Haye, X.~Xue, A.~M. Weiner, and
  R.~Morandotti, ``Micro-combs: {A} novel generation of optical sources,'' {\em
  Physics Reports}, vol.~729, pp.~1--81, Jan. 2018.

\bibitem{Karpov:2019cs}
M.~Karpov, M.~H.~P. Pfeiffer, H.~Guo, W.~Weng, J.~Liu, and T.~J. Kippenberg,
  ``{Dynamics of soliton crystals in optical microresonators},'' {\em Nature
  Physics}, pp.~1--8, Sept. 2019.

\bibitem{Chembo:2010cb}
Y.~K. Chembo and N.~Yu, ``{Modal expansion approach to optical-frequency-comb
  generation with monolithic whispering-gallery-mode resonators},'' {\em Phys.
  Rev. A}, vol.~82, p.~033801, Sept. 2010.

\bibitem{MatskoAB:2011ws}
{Matsko, AB}, A.~A. Savchenkov, W.~Liang, V.~S. Ilchenko, D.~Seidel, and
  L.~Maleki, ``{Mode-locked Kerr frequency combs},'' {\em Opt. Lett.}, vol.~36,
  no.~15, pp.~2845--2847, 2011.

\bibitem{Chembo:2013ew}
Y.~K. Chembo and C.~R. Menyuk, ``{Spatiotemporal Lugiato-Lefever formalism for
  Kerr-comb generation in whispering-gallery-mode resonators},'' {\em Phys.
  Rev. A}, vol.~87, p.~053852, May 2013.

\bibitem{ParraRivas:2014kk}
P.~Parra-Rivas, D.~Gomila, M.~A. Mat{\'\i}as, S.~Coen, and L.~Gelens,
  ``{Dynamics of localized and patterned structures in the Lugiato-Lefever
  equation determine the stability and shape of optical frequency combs},''
  {\em Phys. Rev. A}, vol.~89, p.~043813, Apr. 2014.

\bibitem{Godey:2014cj}
C.~Godey, I.~V. Balakireva, A.~Coillet, and Y.~K. Chembo, ``{Stability analysis
  of the spatiotemporal Lugiato-Lefever model for Kerr optical frequency combs
  in the anomalous and normal dispersion regimes},'' {\em Phys. Rev. A},
  vol.~89, p.~063814, June 2014.

\bibitem{Bao:2017jz}
C.~Bao, H.~Taheri, L.~Zhang, A.~Matsko, Y.~Yan, P.~Liao, L.~Maleki, and A.~E.
  Willner, ``{High-order dispersion in Kerr comb oscillators},'' {\em J. Opt.
  Soc. Am. B}, vol.~34, no.~4, p.~715, 2017.

\bibitem{Godey:2017kt}
C.~Godey, ``{A bifurcation analysis for the Lugiato-Lefever equation},'' {\em
  Eur. Phys. J. D}, vol.~71, p.~063814, May 2017.

\bibitem{Perinet:2017ct}
N.~P{\'e}rinet, N.~Verschueren, and S.~Coulibaly, ``{Eckhaus instability in the
  Lugiato-Lefever model},'' {\em Eur. Phys. J. D}, vol.~71, p.~401, Sept. 2017.

\bibitem{ParraRivas:2018ie}
P.~Parra-Rivas, D.~Gomila, L.~Gelens, and E.~Knobloch, ``{Bifurcation structure
  of localized states in the Lugiato-Lefever equation with anomalous
  dispersion},'' {\em Phys. Rev. E}, vol.~97, no.~4, 2018.

\bibitem{Hansson:2018ie}
T.~Hansson, P.~Parra-Rivas, M.~Bernard, F.~Leo, L.~Gelens, and S.~Wabnitz,
  ``{Quadratic soliton combs in doubly resonant second-harmonic generation},''
  {\em Opt. Lett.}, vol.~43, no.~24, p.~6033, 2018.

\bibitem{Qi:2017da}
Z.~Qi, G.~D'Aguanno, and C.~R. Menyuk, ``{Nonlinear frequency combs generated
  by cnoidal waves in microring resonators},'' {\em J. Opt. Soc. Am. B},
  vol.~34, no.~4, p.~785, 2017.

\bibitem{Qi:2019fla}
Z.~Qi, S.~Wang, J.~Jaramillo-Villegas, M.~Qi, A.~M. Weiner, G.~D'Aguanno, T.~F.
  Carruthers, and C.~R. Menyuk, ``{Dissipative cnoidal waves (Turing rolls) and
  the soliton limit in microring resonators},'' {\em Optica}, vol.~6, no.~9,
  pp.~1220--1232, 2019.

\bibitem{Kholmyansky:2019cy}
D.~Kholmyansky and O.~Gat, ``{Optimal frequency combs from cnoidal waves in
  Kerr microresonators},'' {\em Phys. Rev. A}, vol.~100, no.~6, 2019.

\bibitem{Milian:2015db}
C.~Mili{\'a}n, A.~V. Gorbach, M.~Taki, A.~V. Yulin, and D.~V. Skryabin,
  ``{Solitons and frequency combs in silica microring resonators: Interplay of
  the Raman and higher-order dispersion effects},'' {\em Phys. Rev. A},
  vol.~92, p.~033851, Sept. 2015.

\bibitem{Kartashov:2017cj}
Y.~V. Kartashov, O.~Alexander, and D.~V. Skryabin, ``{Multistability and
  coexisting soliton combs in ring resonators: the Lugiato-Lefever approach},''
  {\em Opt. Express}, vol.~25, no.~10, p.~11550, 2017.

\bibitem{Barashenkov:1996te}
I.~V. Barashenkov and Y.~S. Smirnov, ``{Existence and stability chart for the
  ac-driven, damped nonlinear Schrodinger solitons},'' {\em Phys. Rev. E},
  vol.~54, pp.~5707--5725, Nov. 1996.

\bibitem{Barashenkov:1998wo}
I.~V. Barashenkov, Y.~S. Smirnov, and N.~V. Alexeeva, ``{Bifurcation to
  multisoliton complexes in the ac-driven, damped nonlinear Schrodinger
  equation},'' {\em Phys. Rev. E}, vol.~57, pp.~2350--2364, Feb. 1998.

\bibitem{joshiThermallyControlledComb2016}
C.~Joshi, J.~K. Jang, K.~Luke, X.~Ji, S.~A. Miller, A.~Klenner, Y.~Okawachi,
  M.~Lipson, and A.~L. Gaeta, ``Thermally controlled comb generation and
  soliton modelocking in microresonators,'' {\em Optics Letters}, vol.~41,
  p.~2565, June 2016.

\bibitem{moilleKerrMicroresonatorSolitonFrequency2019}
G.~Moille, X.~Lu, A.~Rao, Q.~Li, D.~A. Westly, L.~Ranzani, S.~B. Papp,
  M.~Soltani, and K.~Srinivasan, ``Kerr-{{Microresonator Soliton Frequency
  Combs}} at {{Cryogenic Temperatures}},'' {\em Physical Review Applied},
  vol.~12, p.~034057, Sept. 2019.

\bibitem{ADeterministicMeth:2020tk}
Z.~Qi, J.~Jaramillo-Villegas, G.~D'Aguanno, T.~F. Carruthers, O.~Gat, A.~M.
  Weiner, and C.~R. Menyuk, ``{A Deterministic Method for Obtaining
  Large-Bandwidth Frequency Combs in Microresonators with Thermal Effects},''
  in {\em Conference on Lasers and Electro-Optics}, 2020.

\bibitem{Guo:2016es}
H.~Guo, M.~Karpov, E.~Lucas, A.~Kordts, M.~H.~P. Pfeiffer, V.~Brasch,
  G.~Lihachev, V.~E. Lobanov, M.~L. Gorodetsky, and T.~J. Kippenberg,
  ``{Universal dynamics and deterministic switching of dissipative Kerr
  solitons in optical microresonators},'' {\em Nature Physics}, vol.~13,
  pp.~94--102, Sept. 2016.

\bibitem{Stone:2018jr}
J.~R. Stone, T.~C. Briles, T.~E. Drake, D.~T. Spencer, D.~R. Carlson, S.~A.
  Diddams, and S.~B. Papp, ``{Thermal and Nonlinear Dissipative-Soliton
  Dynamics in Kerr-Microresonator Frequency Combs},'' {\em Phys. Rev. Lett.},
  vol.~121, no.~6, 2018.

\bibitem{Jiang:2020ia}
X.~Jiang and L.~Yang, ``{Optothermal dynamics in whispering-gallery
  microresonators},'' {\em Light: Science {\&} Applications}, pp.~1--15, Feb.
  2020.

\bibitem{Wang:2014ki}
S.~Wang, A.~Docherty, B.~S. Marks, and C.~R. Menyuk, ``{Boundary tracking
  algorithms for determining the stability of mode-locked pulses},'' {\em J.
  Opt. Soc. Am. B}, vol.~31, no.~11, p.~2914, 2014.

\bibitem{Wang:2016jz}
S.~Wang, B.~S. Marks, and C.~R. Menyuk, ``{Comparison of models of fast
  saturable absorption in passively modelocked lasers},'' {\em Opt. Express},
  vol.~24, no.~18, p.~20228, 2016.

\bibitem{Wang:2016un}
S.~Wang, C.~R. Menyuk, S.~Droste, and L.~Sinclair, ``{Wake Mode Sidebands and
  Instability in Comb Lasers with Slow Saturable Absorbers},'' in {\em 2016
  Conference on Lasers and Electrooptics}, 2016.

\bibitem{Wang:2018gp}
S.~Wang, T.~F. Carruthers, and C.~R. Menyuk, ``{Efficiently modeling the noise
  performance of short-pulse lasers with a computational implementation of
  dynamical methods},'' {\em J. Opt. Soc. Am. B}, vol.~35, no.~10, p.~2521,
  2018.

\bibitem{coleSolitonCrystalsKerr2017}
D.~C. Cole, E.~S. Lamb, P.~Del'Haye, S.~A. Diddams, and S.~B. Papp, ``Soliton
  crystals in {{Kerr}} resonators,'' {\em Nature Photonics}, vol.~11,
  pp.~671--676, Oct. 2017.

\bibitem{hePerfectSolitonCrystals2020}
Y.~He, J.~Ling, M.~Li, and Q.~Lin, ``Perfect {{Soliton Crystals}} on
  {{Demand}},'' {\em Laser \& Photonics Reviews}, vol.~14, p.~1900339, Aug.
  2020.

\bibitem{yiSolitonFrequencyComb2015}
X.~Yi, Q.-F. Yang, K.~Y. Yang, M.-G. Suh, and K.~Vahala, ``Soliton frequency
  comb at microwave rates in a high-{{Q}} silica microresonator,'' {\em
  Optica}, vol.~2, p.~1078, Dec. 2015.

\bibitem{carmonDynamicalThermalBehavior2004}
T.~Carmon, L.~Yang, and K.~J. Vahala, ``Dynamical thermal behavior and thermal
  selfstability of microcavities,'' {\em Optics Express}, vol.~12, no.~20,
  p.~4742, 2004.

\bibitem{obrzudTemporalSolitonsMicroresonators2017}
E.~Obrzud, S.~Lecomte, and T.~Herr, ``Temporal solitons in microresonators
  driven by optical pulses,'' {\em Nature Photonics}, vol.~11, pp.~600--607,
  Sept. 2017.

\bibitem{coleTheoryKerrFrequency2018}
D.~C. Cole, A.~Gatti, S.~B. Papp, F.~Prati, and L.~Lugiato, ``Theory of
  {{Kerr}} frequency combs in {{Fabry}}-{{Perot}} resonators,'' {\em Physical
  Review A}, vol.~98, p.~013831, July 2018.

\bibitem{He:2009fv}
L.~He, Y.-F. Xiao, J.~Zhu, {\c S}.~K. {\"O}zdemir, and L.~Yang, ``{Oscillatory
  thermal dynamics in high-Q PDMS-coated silica toroidal microresonators},''
  {\em Opt. Express}, vol.~17, no.~12, pp.~9571--9581, 2009.

\bibitem{Diallo:2015ck}
S.~Diallo, G.~Lin, and Y.~K. Chembo, ``{Giant thermo-optical relaxation
  oscillations in millimeter-size whispering gallery mode disk resonators},''
  {\em Opt. Lett.}, vol.~40, no.~16, p.~3834, 2015.

\bibitem{Fomin:2005vv}
A.~E. Fomin, M.~L. Gorodetsky, I.~S. Grudinin, and V.~S. Ilchenko,
  ``{Nonstationary nonlinear effects in optical microspheres},'' {\em Journal
  of the Optical Society of America B-Optical Physics}, vol.~22, pp.~459--465,
  Feb. 2005.

\bibitem{Park:2007wf}
Y.-S. Park and H.~Wang, ``{Regenerative pulsation in silica microspheres},''
  {\em Opt. Lett.}, vol.~32, no.~21, pp.~3104--3106, 2007.

\bibitem{baoDirectSolitonGeneration2017}
C.~Bao, Y.~Xuan, J.~A. {Jaramillo-Villegas}, D.~E. Leaird, M.~Qi, and A.~M.
  Weiner, ``Direct soliton generation in microresonators,'' {\em Optics
  Letters}, vol.~42, pp.~2519--2522, July 2017.

\bibitem{Strogatz:1994tz}
S.~H. Strogatz, {\em {Nonlinear dynamics and Chaos: with applications to
  physics, biology, chemistry, and engineering}}.
\newblock Studies in nonlinearity, Addison Wesley, 1994.

\end{thebibliography}
\end{document}